# Optimal Consumption and Investment with Fixed and Proportional Transaction Costs [*]

Albert Altarovici[†]    Max Reppen[†]    H. Mete Soner[†]

October 13, 2016

### Abstract

The classical optimal investment and consumption problem with infinite horizon is studied in the presence of transaction costs. Both proportional and fixed costs as well as general utility functions are considered. Weak dynamic programming is proved in the general setting and a comparison result for possibly discontinuous viscosity solutions of the dynamic programming equation is provided. Detailed numerical experiments illustrate several properties of the optimal investment strategies.

## 1 Introduction

In this paper, we study the Merton problem with fixed and proportional transaction costs. The problem introduced and solved by Merton [34] is now a classical application of stochastic control to mathematical finance and paved the way for decades of research on extensions to more general settings. The problem, simply stated, is to determine the optimal investment strategy for a rational, risk-averse agent in a market consisting of one risky asset (stock) and one riskless asset (bank account). The latter grows at the risk-free rate, whereas the former has a higher mean rate of return but is subject to risk in the form of volatility. Merton found that the optimal strategy was to maintain a constant proportion of total wealth in the risky asset. That amount, called the Merton proportion, would depend on the market parameters as well as the investor's risk preferences. Since even the simplest models of a stock price assumes a (geometric) Brownian motion driver, the investor would have to continuously rebalance his/her portfolio in order to implement the constant proportion strategy and achieve optimal returns. However, in the presence of transaction costs following such a strategy would immediately lead to bankruptcy. Thus the question arises: *what is the optimal strategy when there are transaction costs?*

---

[*]Research is partly supported by the ETH Foundation, the Swiss Finance Institute and a Swiss National Foundation grant SNF 200021_153555.

[†]Departement für Mathematik, Rämistrasse 101, CH-8092, Zürich, Switzerland

Arguably, the most well-known example of a transaction cost that arises in practice is the *bid-ask spread*. We often refer to this cost as a *proportional* cost since it accrues in proportion to the size of the trade. The first to study the Merton problem with proportional costs were Constantinides and Magill [32] and later again Constantinides in [9]. In these works, the notion of a buffer region around the Merton optimal proportion within which the investor does not re-balance his/her portfolio was first introduced. While it was not clear at the time what the optimal strategy would have to be, they were nonetheless able to argue that transaction costs had a profound negative impact on investment returns. Soon after, the insights provided by these works were further developed [13, 14, 44] to not only determine the form of the optimal strategies, but help place the problem on firm theoretical footing. The optimal strategy in this case is as follows: do not re-balance while the portfolio is contained inside the buffer region. However, once the boundary of the region is breached, transact minimally so as to remain in the region.

Proportional transaction costs have received most of the attention in the literature for a number of reasons. First, they are relevant to investors of all sizes. Second, from a mathematical standpoint at least, their scale invariance is a useful property. Third, although a lot of work has been done, there are still plenty of questions left to answer. Arguably, the most important of which is: *how does one compute the buffer region?*

Partial answers to this question are provided in restricted settings. Davis and Norman [13] showed that trading boundaries can be determined numerically by solving a free-boundary problem. In the asymptotic limit for small costs, the no-trade region and the corresponding utility loss can be determined explicitly at the leading order, cf. Shreve and Soner [44], Whalley and Wilmott [48], Janeček and Shreve [22], as well as many more recent studies [6, 18, 47, 38, 7]. Extensions to more general preferences and stochastic opportunity sets have been studied numerically by Balduzzi, Lynch, and Tan [30, 3, 31]. Corresponding formal asymptotics have been determined by Goodman and Ostrov [19], Martin [33], Kallsen and Muhle-Karbe [25, 24] as well as Soner and Touzi [47]. The last study, [47], also contains a rigorous convergence proof for general utilities, which is extended to several risky assets by Possamaï, Soner, and Touzi [38].

It has been determined that proportional costs lead to strategies which rely on infinitely many small (local-time) transactions to produce optimal outcomes. However, this theoretical result is unsatisfying from a practical point of view; conducting infinitely many trades in any finite time horizon is impossible. On the other hand, the inclusion of fixed costs (e.g. a brokerage fee of $1 paid each time the investor trades) only allows for a finite number of trades over finite time intervals. In this case using the class of strategies which are optimal for proportional costs leads to immediate bankruptcy in the face of fixed costs (cf. [43]). When both costs are present, the optimal strategy prescribes using two (inner and outer) buffer regions and re-balancing happens as follows: the investor is inactive when the portfolio is inside the outer region. Once the outer region is breached, he/she trades to the boundary of the inner region. These cost structures lead to strategies that are more appealing from a practical standpoint;

now the investor must determine (a set of stopping) times at which it is optimal to transact. The impulse control problem in the context of portfolio management was first approached by [15].

The main drawback to fixed costs from a modeling point of view is that they destroy the favorable scaling properties that usually allow one to reduce the dimensionality of the problem for utilities with constant relative or absolute risk aversion. In particular, the no-trade region is no longer a cone, even in the simplest settings with constant investment opportunities as well as constant absolute or relative risk aversion. Accordingly, the literature analyzing the impact of fixed trading costs is much more limited than for proportional costs: on the one hand, there are a number of numerical studies [42, 28], which iteratively solve the dynamic programming equations. On the other hand, Korn [26] as well as Lo, Mamaysky, and Wang [29] have obtained formal asymptotic results for investors with constant *absolute* risk aversion. This structure resembles that of inventory problems where both fixed and proportional costs are present. Scarf [41] introduced the notion of $K$-convexity to analyze these problems in one space dimension, which was later successfully used in [39] and [5].

For small costs, these authors find that constant trading boundaries are optimal at the leading order. Thus, these models are tractable but do not allow us to study how the impact of fixed trading costs depends on the size of the investor under consideration. The same applies to the "quasi-fixed" costs proposed by Morton and Pliska [35], and analyzed in the small-cost limit by Atkinson and Wilmott [2]. In their model, each trade – regardless of its size – incurs a cost proportional to the investors' current wealth, leading to a scale-invariant model where investors of all sizes are affected by the "quasi-fixed" costs to the same extent. The fixed (in the sense of Morton and Pliska) and proportional cost problem was analyzed by Irle and Sass [21, 20] in the context of maximizing the asymptotic growth rate of the portfolio value. They show that "constant-boundary" policies are optimal controls; these strategies dictate that there is an inner and an outer threshold around the optimal frictionless portfolio allocation in which the agent trades to the inner threshold when the outer one is breached. Korn [26] considers the same transaction cost structure as we do, i.e. fixed and proportional costs where the fixed costs are independent of the investor's current wealth. For the formulation and the analysis of multi-dimensional problems we refer the reader to the book by Kabanov and Safarian [23].

Shreve and Soner [44] were the first to provide a rigorous analysis of the value function in the frictional (proportional cost) setting using the tools of viscosity theory. In the mixed cost (fixed and proportional) setting, Øksendal and Sulem [37] study the case of one risky asset and power utility with risk aversion between 0 and 1. Under some technical assumptions, they prove that the value function is a viscosity solution of the associated Hamilton-Jacobi-Bellman inequality and also provide a comparison result in that case. More recently, Belak, Menkens, and Sass [4] revisit the comparison problem in the case of proportional transaction costs and they prove uniqueness in the case of power and logarithmic utilities. However, the question of uniqueness for more general utilities and the analysis of higher dimensions and the mixed case of proportional and fixed costs remained open. We address all these issues.

Convex duality is also used to analyse the more general semi-martingale models. We refer the reader to the initial paper of Cvitanic and Karatzas [11], the recent manuscript of Czichowsky and Schachermayer [12], and the references therein.

## Main themes

The results contained in this paper consist of two themes. The first is the analysis of the value function for the general impulse control problem in multi-dimensions, assuming constant coefficient, correlated geometric Brownian motion stock price dynamics and general utilities with asymptotic elasticity less than 1. We do not assume *a priori* measurability of the value function and we prove in Section 3.1 that it satisfies a weak dynamic programming principle (DPP). Using the weak DPP, in Section 3.2 we prove that the value function is a constrained viscosity solution of the associated quasi-variational Hamilton-Jacobi-Bellman equation (dynamic programming equation, or DPE for short). Finally, we prove a comparison principle, Theorem 4.3, for the DPE which circumvents the boundary singularities associated with general utility functions. This result is proved for general discontinuous sub and super solutions. The main contribution here is a novel technique of an appropriate state-space translation of the viscosity super-solution in order to avoid the boundary issues. Moreover, it provides a uniqueness result, Theorem 4.4, for possibly discontinuous solutions. Since the standard comparison would imply the continuity of the unique solution, the statements of both Theorem 4.3 and Theorem 4.4 are non-standard and novel.

The second theme is to determine the shape of the asymptotic no-trade region when there is more than one risky asset in the presence of mixed transaction costs and/or multiple fixed costs. This includes a formal derivation of the corrector equations obtained by homogenization of the DPE. These, in turn, correspond to the DPE of an ergodic impulse control problem. This control problem seems to only admit explicit solutions in either 1-D or when there are no proportional costs. Nevertheless, we are able to calculate the optimal controls and no-trade regions numerically using a policy iteration scheme. By appropriately modifying the control penalization structure, we are able to investigate a diverse set of transaction cost structures. We perform benchmark tests on the numerical scheme using our knowledge of the explicit solutions of the 1-D models and the fixed cost solution in 2-D.

## 2  Model and main results

### 2.1  Market, trading strategies, and wealth dynamics

Consider a financial market consisting of a safe asset earning a constant interest rate $r > 0$, and $d$ risky assets with expected excess returns $\mu^i - r > 0$ and invertible infinitesimal covariance matrix $\sigma\sigma^\top$:

$$dS_t^0 = S_t^0 r dt, \quad dS_t = S_t \mu dt + S_t \sigma dW_t,$$

for a $d$-dimensional standard Brownian motion $(W_t)_{t\geq 0}$ defined on a filtered probability space $(\Omega, \mathscr{F}, (\mathscr{F}_t)_{t\geq 0}, P)$, where $(\mathscr{F}_t)_{t\geq 0}$ denotes the augmentation of the filtration generated by $(W_t)_{t\geq 0}$. Each trade incurs a *fixed transaction cost* $\lambda_f > 0$ and a *proportional transaction cost* $\lambda_p \geq 0$. As a result, portfolios can only be rebalanced finitely many times over finite time intervals, and trading strategies can be described by pairs $(\tau, m)$, where the trading times $\tau = (\tau_1, \tau_2, \ldots)$ are a sequence of stopping times (strictly) increasing towards infinity, and the $\mathscr{F}_{\tau_i}$-measurable, $\mathbb{R}^d$-valued random variables collected in $m = (m_1, m_2, \ldots)$ describe the transfers at each trading time. More specifically, $m_i^j$ represents the monetary amount transferred from the safe to the $j$-th risky asset at time $\tau_i$. Each trade is assumed to be self-financing, and the costs are deducted from the safe asset account. Thus, the safe and risky positions evolve as

$$(x,y) = (x, y^1, \ldots, y^d) \mapsto \left(x - \sum_{j=1}^d (1+\lambda_p \mathrm{sgn}(m_i^j))m_i^j - \lambda_f, y^1 + m_i^1, \ldots, y^d + m_i^d\right)$$

for each trade $m_i$ at time $\tau_i$. The investor also consumes from the safe account at some rate $(c_t)_{t\geq 0}$. Hence, the wealth dynamics corresponding to a *consumption-investment strategy* $\nu = (c, \tau, m)$ starting from an initial position $(X_{0-}, Y_{0-}) = (x, y) \in \mathbb{R} \times \mathbb{R}^d$, are given by

$$X_t = x + \int_0^t (rX_s - c_s) ds - \sum_{k=1}^\infty \left(\lambda_f + \sum_{j=1}^d (1+\lambda_p \mathrm{sgn}(m_k^j))m_k^j\right) 1_{\{\tau_k \leq t\}},$$

$$Y_t^i = y^i + \int_0^t Y_s^i \frac{dS_s^i}{S_s^i} + \sum_{k=1}^\infty m_k^i 1_{\{\tau_k \leq t\}}.$$

We write $(X, Y)^{\nu, x, y}$ for the solution of the above equation. The *solvency region*

$$\mathrm{K}_\lambda := \left\{(x, y) \in \mathbb{R}^{d+1} : \max\left\{ x + y \cdot \mathbf{1}_d - \lambda_p \|y\|_1 - \lambda_f \, , \, \min_{i=1,\ldots,d}\{x, y^i\} \right\} \geq 0\right\}$$

is the set of positions with nonnegative liquidation value. Here, $\mathbf{1}_d = (1, \ldots, 1) \in \mathbb{R}^d$ and $\|\cdot\|_1$ denotes the *Manhattan distance*, namely $\|v\|_1 := \sum_{i=1}^d |v^i|$. A visualization of this set in the case of one risky asset is given in Figure 1.

A strategy $\nu = (c, \tau, m)$ starting from the initial position $(x, y)$ is called *admissible* if it remains solvent at all times: $(X, Y)_t^{\nu, x, y} \in \mathrm{K}_\lambda$, for all $t \geq 0$, $P$-a.s. The set of all admissible strategies is denoted by $\Theta^\lambda(x, y)$.

## 2.2 Preferences

In the above market with constant investment opportunities $(r, \mu, \sigma)$ and transaction costs $\lambda = (\lambda_f, \lambda_p) \in \mathbb{R}_{>0} \times \mathbb{R}_{\geq 0}$, an investor with utility function $U$ and

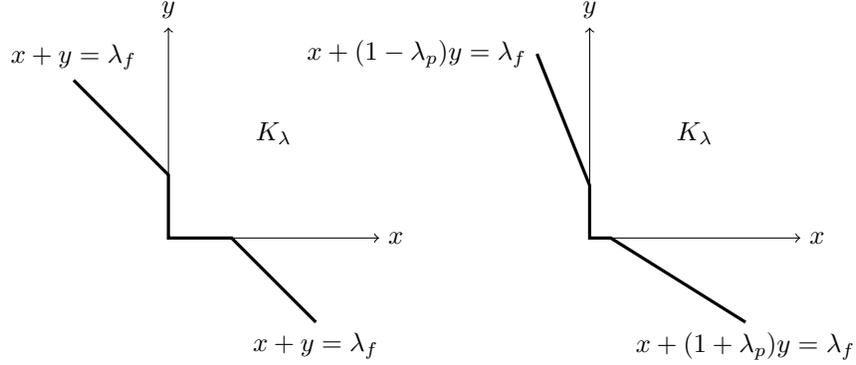

Figure 1: With only one risky asset the solvency region is bounded by four lines, here drawn thick. Note that in the left figure, where $\lambda_p = 0$, the solvency region is unbounded for every fixed $z = x + y$, causing some extra obstacles in the analysis of this special case. Note that any trade move the position parallel to one of the two sloping lines, shifting inwards by the size of $\lambda_f$.

*impatience rate* $\beta > 0$ trades to maximize the expected utility from consumption over an infinite horizon, starting from an initial endowment of $X_{0-} = x$ in the safe and $Y_{0-} = y$ in the risky assets, respectively. The utility function $U : [0, \infty) \to \{-\infty\} \cup \mathbb{R}$ is real-valued, smooth, increasing, and strictly concave on $(0, \infty)$ and has asymptotic elasticity $1 - \gamma$. We assume that $U(0) = U(0^+)$.

The value of the investor's consumption is given by[1]

$$v^\lambda(x, y) = \sup_{(c,\tau,m) \in \Theta^\lambda(x,y)} \mathbb{E}\left[\int_0^\infty e^{-\beta t} U(c_t) dt\right]. \quad (2.1)$$

The expected qualitative properties of an optimal strategy with one risky asset are sketched in Figure 2.

**Assumption 2.1.** *We shall assume throughout that*

$$\frac{\beta}{\gamma} + \left(1 - \frac{1}{\gamma}\right)\left(r + \frac{(\mu - r\mathbf{1}_d)^\top (\sigma^\top \sigma)^{-1}(\mu - r\mathbf{1}_d)}{2\gamma}\right) > 0.$$

**Remark 2.2.** *This assumption equivalent to saying that the frictionless value function is finite in the case where $U$ is a power (or logarithmic) utility function with risk-aversion parameter $\gamma$.*

*In fact, if $U$ is a more general utility function with asymptotic elasticity $1 - \gamma < 1$, the value function is still finite. More precisely, for all $0 < \hat{\gamma} < \gamma$ such that the assumption holds with $\gamma$ replaced by $\hat{\gamma}$, there exist $C_1$ and $C_2$ such that*

$$v^\lambda(x, y) \leq C_1(1 + v(x, y)) \leq C_2(1 + |x + y \cdot \mathbf{1}_d|^{1-\hat{\gamma}}),$$

---

[1] By convention, the value of the integral is set to minus infinity if its negative part is infinite.

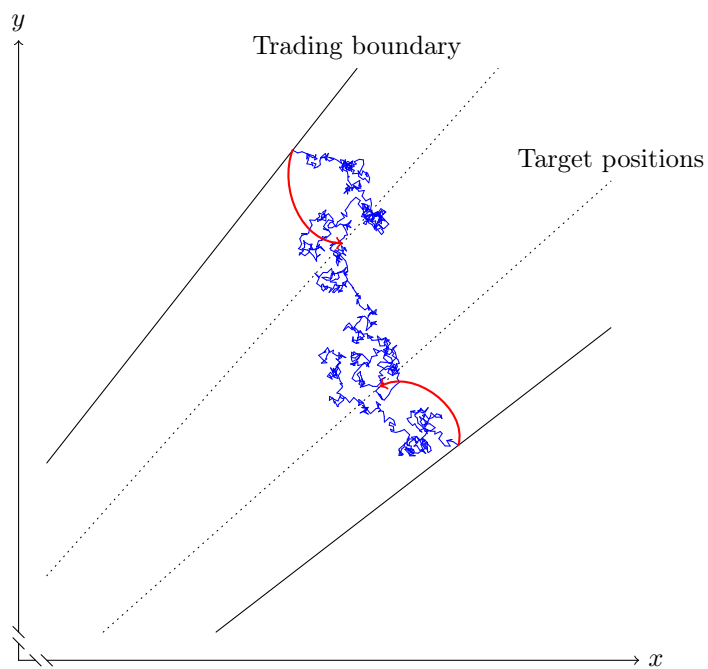

Figure 2: Sample path of portfolio positions. The trader stays passive as long as the portfolio remains in some region, and trades inwards to a target line when the process crosses the trading boundary. The blue lines indicate movement of the portfolio from price changes, whereas the red arcs indicate trades.

*where $v$ here denotes the frictionless value function with power (or logarithmic) utility function and risk-aversion parameter $\hat{\gamma}$. This is a direct consequence of the properties of $U$ proven in [27].*

We study the value function on its effective domain

$$\mathfrak{D} := \{(x,y) \in \mathrm{K}_\lambda \ : \ |v^\lambda(x,y)| \neq \infty\}.$$

Our first result shows that it is a *constrained viscosity solution* of the Dynamic Programming Equation (2.4) on its effective domain $\mathfrak{D}$ as defined in [45, 46]; see also the Definition 3.4, below.

Note that $\mathrm{int}(\mathrm{K}_\lambda) \subset \mathfrak{D}$. This is clear by Remark 2.2 and the observation that for any $(x,y) \in \mathrm{int}(\mathrm{K}_\lambda)$ such that a ball centered in $(x,y)$ with radius $\delta$ is contained in $\mathrm{K}_\lambda$, it is admissible to liquidate with at least $\delta$ cash left and thereafter consume the interest. This yields

$$v^\lambda(x,y) \geq \frac{U(r\delta)}{\beta} > -\infty.$$

More precisely, the following hold:

(i) If $U(0) > -\infty$, then $\mathrm{K}_\lambda = \mathfrak{D}$.

(ii) If $U(0^+) = -\infty$, then $\{(x,y^1,\ldots,y^d) \in \mathbb{R}^{d+1}_{\geq 0} \ : \ x > 0\} \cup \mathrm{int}(\mathrm{K}_\lambda) = \mathfrak{D}$.

**Theorem 2.3.** *Under Assumption 2.1 the value function $v^\lambda$ of the Merton problem with fixed costs $\lambda_f > 0$ and proportional costs $\lambda_p \geq 0$ is a (possibly) discontinuous constrained viscosity solution of the Dynamic Programming Equation (2.4) on its effective domain $\mathfrak{D}$.*

**Remark 2.4.** *The value function is indeed discontinuous in $\mathrm{K}_\lambda$, and in the case where $U(0) > -\infty$ this is clearly true also in $\mathfrak{D}$. In the corner $(\lambda_f, \mathbf{0}_d)$ of $\partial \mathrm{K}_\lambda$ the investor may decide to not trade and consume the interest rate, thus obtaining the payoff $U(r\lambda_f)/\beta$. On the other hand, for any point $(x,y) \in \partial \mathrm{K}_\lambda \setminus \mathbb{R}^{d+1}_{\geq 0}$, the only admissible strategy is to liquidate to $(0, \mathbf{0}_d)$. In such cases we must have $v^\lambda(x,y) = U(0)/\beta$. Hence, no matter how close $(x,y)$ is to $(\lambda_f, \mathbf{0}_d)$,*

$$v^\lambda(\lambda_f, \mathbf{0}_d) - v^\lambda(x,y) \geq \frac{U(r\lambda_f) - U(0)}{\beta}.$$

*This value is strictly greater than 0 and independent of the choice of point. Therefore, $v^\lambda$ has a discontinuity at the point $(\lambda_f, \mathbf{0}_d)$.*

### 2.3 The frictional dynamic programming equation

For the convenience of the reader, we now recall how to heuristically derive the dynamic programming equation with fixed trading costs. We start from the ansatz that the value function $v^\lambda(x,y)$ for our infinite horizon problem with constant model parameters should only depend on the positions in each of the assets. Evaluated along the positions $X_t, Y_t$ corresponding to any admissible policy $\nu = (c, \tau, m)$, Itô's formula in turn yields

$$dv^\lambda(X_t, Y_t) \tag{2.2}$$
$$= \Big(v_x^\lambda(X_t, Y_t)(rX_t - c_t) + \mu \cdot \mathbf{D}_y v^\lambda(X_t, Y_t) + \frac{1}{2}\text{Tr}[\sigma\sigma^\top \mathbf{D}_{yy} v^\lambda(X_t, Y_t)]\Big)dt$$
$$+ \mathbf{D}_y v^\lambda(X_t, Y_t)^\top \sigma dW_t$$
$$+ \sum_{\tau_i \leq t} \big(v^\lambda(X_{\tau_i} - m_i \cdot \mathbf{1}_d - \lambda, Y_{\tau_i} + m_i) - v^\lambda(X_{\tau_i}, Y_{\tau_i})\big),$$

where

$$\mathbf{D}_y^i = y^i \frac{\partial}{\partial y_i}, \quad \mathbf{D}_{yy}^{ij} = y^i y^j \frac{\partial^2}{\partial y^i \partial y^j}, \quad i, j = 1, \ldots, d.$$

By the martingale optimality principle of stochastic control, the utility

$$\int_0^t e^{-\beta s} U(c_s) ds + e^{-\beta t} v^\lambda(X_t, Y_t)$$

obtained by applying an arbitrary policy $\nu$ until some intermediate time $t$ and then trading optimally should always lead to a supermartingale, and to a martingale if the optimizer is used all along. Between trades – in the policy's "no-trade region" – this means that the absolutely continuous drift should be nonpositive, and zero for the optimizer. After taking into account (2.2), using integration by parts, and canceling the common factor $e^{-\beta t}$, this leads to

$$0 = \sup_{c>0} \Big\{ -\beta v^\lambda(x, y) + U(c) + (rx - c)v_x^\lambda(x, y) \tag{2.3}$$
$$+ \mu \cdot \mathbf{D}_y v^\lambda(x, y) + \frac{1}{2}\text{Tr}[\sigma\sigma^\top \mathbf{D}_{yy}] v^\lambda(x, y)\Big\}.$$

By definition, the value function can only be decreased by admissible bulk trades at any time:

$$0 \geq \sup_{m \in \mathbb{R}^d} \{v^\lambda(x - m \cdot \mathbf{1}_d - \|m\|_1 \lambda_p - \lambda_f, y + m) - v^\lambda(x, y)\},$$

and this inequality should become an equality for the optimal transaction once the boundaries of the no-trade region are breached. Combining this with (2.3) and switching the sign yields the *dynamic programming equation*:

$$0 = \min\{\beta v^\lambda - \widetilde{U}(v_x^\lambda) - \mathscr{L} v^\lambda, v^\lambda - \mathbf{M} v^\lambda\}, \tag{2.4}$$

where $\widetilde{U}(\tilde{c}) = \sup_{c>0}(U(c) - c\tilde{c})$ is the convex dual of the utility function $U$, the differential operator $\mathscr{L}$ is defined as

$$\mathscr{L} = rx \frac{\partial}{\partial x} + \mu \cdot \mathbf{D}_y + \frac{1}{2}\text{Tr}[\sigma\sigma^\top \mathbf{D}_{yy}],$$

and **M** denotes the non-local *intervention operator*

$$\mathbf{M}\psi(x,y) = \sup_{m \in \mathbb{R}^d} \{\psi(x',y') : (x',y') = (x - m \cdot \mathbf{1}_d - \|m\|_1 \lambda_p - \lambda_f, y + m) \in \mathrm{K}_\lambda\}. \tag{2.5}$$

In the event that there is no $m \in \mathbb{R}^d$ for which $(x',y') = (x - m \cdot \mathbf{1}_d - \|m\|_1 \lambda_p - \lambda_f, y + m) \in \mathrm{K}_\lambda$, then $\mathbf{M}\psi(x,y) = -\infty$.

## 3 Proof of Theorem 2.3

In this section we prove that the value function $v^\lambda$ is a constrained viscosity solution of the corresponding Dynamic Programming Equation (2.4) on its effective domain. We present a direct proof of the weak dynamic programming principle. For a more general approach, we refer to [16]. Then, we use it to prove that $v^\lambda$ is indeed a viscosity solution of (2.4).

### 3.1 Weak dynamic programming principle for $v^\lambda$

Fix $(x,y) \in \mathfrak{D}$ and $M := 2(x + y \cdot \mathbf{1}_d)$. Set

$$\mathcal{O}_\lambda := \mathcal{O}_\lambda(x,y;M) = \{(x',y') \in \mathfrak{D} \ : \ x' + y' \cdot \mathbf{1}_d < M\}.$$

Define $B_\delta(x,y) \subset \mathbb{R}^{d+1}$ to be a (relatively) open ball (in $\mathrm{K}_\lambda$) of radius $\delta$ centered at $(x,y)$. Choose $\delta > 0$ sufficiently small so that $\delta < \lambda$, $B_\delta(x,y) \subset \mathcal{O}_\lambda$, and $(0,0) \notin \overline{B_\delta(x,y)}$. For any investment-consumption policy $\nu$ and initial endowment $(x',y') \in B_\delta(x,y)$, define $\theta := \theta^\nu$ as the exit time of the state process $(X,Y)^{\nu,x',y'}$ from $B_\delta(x,y)$. Following standard convention, our notation does not explicitly show the dependence of $\theta$ on $\nu$. It is then clear that

$$(X_{\theta^-}, Y_{\theta^-}) \in \overline{B_\delta(x,y)} \quad \text{and} \quad (X_\theta, Y_\theta) \in \mathcal{O}_\lambda.$$

Let $\varphi$ be a bounded,[2] upper-semicontinuous function on $\mathcal{O}_\lambda$ which is locally $C^2$ at $(x,y)$ and satisfies

$$v^\lambda \leq \varphi \quad \text{on } \mathcal{O}_\lambda.$$

Without loss of generality, we assume that $\delta > 0$ was chosen small enough that $\varphi \in C^2(B_\delta(x,y))$. Then, we have

$$v^\lambda(x,y) \leq \sup_{\nu \in \Theta^\lambda(x,y)} \mathbb{E}\left[\int_0^\theta e^{-\beta t} U(c_t) dt + e^{-\beta \theta} \varphi(X_\theta, Y_\theta)\right]. \tag{3.1}$$

Conversely, let $\varphi$ be a smooth function on $\mathcal{O}_\lambda$, satisfying

$$v^\lambda \geq \varphi \quad \text{on } \mathcal{O}_\lambda.$$

---
[2] We can take bounded test functions because $v^\lambda$ is bounded from above on $\mathcal{O}_\lambda$.

Then, we have

$$v^\lambda(x,y) \geq \sup_{\nu \in \Theta^\lambda(x,y)} \mathbb{E}\left[\int_0^\theta e^{-\beta t}U(c_t)dt + e^{-\beta\theta}\varphi(X_\theta, Y_\theta)\right]. \qquad (3.2)$$

Without loss of generality, let $\Omega = C_0([0,\infty), \mathbb{R}^d)$ be the space of continuous function starting at zero, equipped with the the Wiener measure $\mathbb{P}$, a standard Brownian motion $W$, and the completion $\{\mathcal{F}_t\}_{t\geq 0}$ of the filtration generated by $W$.

Given a control $\nu \in \Theta^\lambda(x,y)$ and the exit time $\theta := \theta^\nu$ from above, fix $\omega \in \Omega$ and define

$$\nu^{\theta,\omega}(\omega', t) := \nu\big(\omega \overset{\theta}{\oplus} \omega', t + \theta(\omega)\big), \qquad \forall \omega' \in \Omega, t \geq 0,$$

where

$$(\omega \overset{\theta}{\oplus} \omega')_t = \begin{cases} \omega_t & \text{if } t \in [0, \theta(\omega)) \\ \omega'_{t-\theta(\omega)} + \omega_{\theta(\omega)} & \text{if } t \geq \theta(\omega). \end{cases}$$

We start with the proof of (3.1). By construction,

$$\nu^{\theta,\omega} \in \Theta^\lambda((X_{\theta(\omega)}, Y_{\theta(\omega)})^{\nu,x,y});$$

in particular, $\nu^{\theta,\omega}$ is a well-defined impulse control. Therefore,

$$\mathbb{E}\left[\int_0^\infty e^{-\beta t}U(c_t^\nu)dt \,\bigg|\, \mathcal{F}_\theta\right](\omega)$$
$$= \int_0^{\theta(\omega)} e^{-\beta t}U\big(c_t^\nu(\omega)\big)dt + e^{-\beta\theta(\omega)} \int_\Omega \int_0^\infty e^{-\beta t}U\big(c_t^{\nu^{\theta,\omega}}(\omega')\big)dt\,d\mathbb{P}(\omega')$$
$$\leq \int_0^{\theta(\omega)} e^{-\beta t}U\big(c_t^\nu(\omega)\big)dt + e^{-\beta\theta(\omega)} v^\lambda((X_{\theta(\omega)}, Y_{\theta(\omega)})^{\nu,x,y})$$
$$\leq \int_0^{\theta(\omega)} e^{-\beta t}U(c_t^\nu(\omega))dt + e^{-\beta\theta(\omega)} \varphi((X_{\theta(\omega)}, Y_{\theta(\omega)})^{\nu,x,y}).$$

As a result, for any $\nu \in \Theta^\lambda(x,y)$:

$$\mathbb{E}\left[\int_0^\infty e^{-\beta t}U(c_t^\nu)dt\right] \leq \mathbb{E}\left[\int_0^\theta e^{-\beta t}U(c_t^\nu)dt + e^{-\beta\theta}\varphi((X_\theta, Y_\theta)^{\nu,x,y})\right].$$

By taking the supremum over all policies $\nu$, we arrive at (3.1).

To prove (3.2), set $\mathbb{V}$ to be the right hand side of (3.2):

$$\mathbb{V} := \sup_{\nu \in \Theta^\lambda(x,y)} \mathbb{E}\left[\int_0^\theta e^{-\beta t}U(c_t^\nu)dt + e^{-\beta\theta}\varphi((X_\theta, Y_\theta)^{\nu,x,y})\right].$$

For any $\eta > 0$, we can choose $\nu^\eta \in \Theta_\lambda(x,y)$ satisfying

$$\mathbb{V} \leq \eta + \mathbb{E}\left[\int_0^\theta e^{-\beta t}U(c_t^{\nu^\eta})dt + e^{-\beta\theta}\varphi((X_\theta, Y_\theta)^{\nu,x,y})\right]. \qquad (3.3)$$

We begin by covering first the interior, $\mathring{\mathcal{O}}_\lambda$. For every point $\zeta = (\tilde{x}, \tilde{y})$ in $\mathring{\mathcal{O}}_\lambda$, set

$$R(\zeta) := R_\eta(\tilde{x}, \tilde{y})$$
$$= \{(x', y') \in \mathring{\mathcal{O}}_\lambda \ : \ x' > \tilde{x}, \ y' > \tilde{y}, \ \varphi(x', y') < \varphi(\tilde{x}, \tilde{y}) + \eta\}.$$

Since $\varphi$ is smooth, each $R(\zeta)$ is open and

$$\mathring{\mathcal{O}}_\lambda \subset \bigcup_{\zeta \in \hat{\mathcal{O}}_\lambda} R(\zeta).$$

Hence, by the Lindelöf covering lemma, we can extract a countable subcover

$$\mathring{\mathcal{O}}_\lambda \subset \bigcup_{n \in \mathbb{N}} R(\zeta_n).$$

Only the boundary, $\partial K_\lambda$, remains to be covered. It is convenient to write

$$\partial K_\lambda = S \cup C,$$

and $\hat{\mathcal{O}}_\lambda := \mathcal{O}_\lambda \setminus S$ where

$$S = \{\mathbf{v} = (v^0, \ldots v^d) \in \partial K_\lambda : \exists i, v^i < 0\} \tag{3.4}$$

and $C$ is the relative complement of $S$ in $\partial K_\lambda$. Note that this means that $C$ is the boundary of a $(d+1)$-simplex bounded by the coordinate axes and the plane $\{v \in \mathbb{R}^{d+1} : v^0 + \sum_{i=1}^d (1 - \lambda_p) v^i = \lambda_f\}$, with the open face in this plane removed.

The interior of each $k$-simplex in $C$, for $0 < k \leq d$, can be written as

$$C_k(I_k) = \left\{(v^0, \ldots, v^d) \in \mathbb{R}^{d+1}_{\geq 0} : v^i > 0 \Leftrightarrow i \in I_k, \sum_{i \in I_k} (1 - \lambda_p 1_{\{i \neq 0\}}) v^i < \lambda_f\right\}$$

for some set of distinct indices $I_k = \{i_j : j = 1, \ldots, k\}$. For each such $k$-simplex, we cover its interior using sets of the form

$$R(\zeta) = \{\hat{\zeta} \in C_k(i_1, \ldots, i_k) \ : \ \hat{\zeta} > \zeta, \ \varphi(\hat{\zeta}) < \varphi(\zeta) + \eta\}.$$

Clearly, each
$$C_k(i_1, \ldots, i_k) \subset \bigcup_{\zeta \in C_k(i_1, \ldots, i_k)} R(\zeta)$$

and we can again extract a countable subcover. We remark that if $U(0+) = -\infty$, there is no need to create a covering of any simplex other the ones with non-trivial first coordinate component (since any other simplex is not contained in the effective domain of $v^\lambda$). If instead, $U(0) > -\infty$, then we need to cover everything. Finally, note that we also do not need to cover $S$ because, provided $\eta$ is sufficiently small, an $\eta$-optimal strategy $\nu^\eta$ will force $(X_\theta, Y_\theta)^{\nu^\eta, x, y} \notin S$.

So far we have created a countable covering $\{R(\zeta_n)\}_{n=d+2}^{\infty}$ (up to re-indexing) of $\hat{\mathcal{O}}_\lambda \setminus \{\lambda_f \mathbf{e}_0, \frac{\lambda_f}{1-\lambda_p}\mathbf{e}_1, \ldots \frac{\lambda_f}{1-\lambda_p}\mathbf{e}_d, \mathbf{0}_{d+1}\}$, where $\mathbf{e}_i$ denotes the $i$-th elementary unit vector in $\mathbb{R}^{d+1}$ with indexing starting at 0. For each $i = 1, \ldots, d$, set $\zeta_i := \frac{\lambda_f}{1-\lambda_p}\mathbf{e}_i$. Then set $\zeta_0 = \lambda_f \mathbf{e}_0$ and $\zeta_{d+1} = \mathbf{0}_{d+1}$. Finally, define $R(\zeta_i) := \{\zeta_i\}$ for $0 \leq i \leq d+1$. Thus we have a countable covering $\{R(\zeta_n)\}_{n=0}^{\infty}$ of $\hat{\mathcal{O}}$.

Now, define a mapping $\mathcal{I} : \hat{\mathcal{O}}_\lambda \to \mathbb{N}$:

$$\mathcal{I}(x', y') := \min\{n : (x', y') \in R(\zeta_n)\}, \quad \forall (x', y') \in \hat{\mathcal{O}}_\lambda$$

and set

$$\zeta(x', y') := \zeta_{\mathcal{I}(x', y')},$$

By definition, these constructions imply

$$\varphi(x', y') \leq \varphi(\zeta(x', y')) + \eta, \quad \forall (x', y') \in \hat{\mathcal{O}}_\lambda. \tag{3.5}$$

For each $n \in \mathbb{N}$, we choose a control $\nu^n \in \Theta^\lambda(\zeta_n)$ so that

$$v^\lambda(\zeta_n) \leq \mathbb{E}\left[\int_0^\infty e^{-\beta t} U(c_t^{\nu^n}) dt\right] + \eta. \tag{3.6}$$

Note that for each $n \geq 0$, $\nu^n \in \Theta^\lambda(x', y')$ for every $(x', y') \in R(\zeta_n)$. We now define a composite strategy $\nu^*$, which follows the policy $\nu^\eta$ satisfying (3.3) until the corresponding state process $(X, Y)^{\nu^\eta, x, y}$ leaves $B_\delta(x, y)$ at time $\theta = \theta^{\nu^\eta}$. We have already argued that $(X_\theta, Y_\theta)^{\nu^\eta, x, y} \in \hat{\mathcal{O}}_\lambda$. The policy thereafter is $\nu^n$ corresponding to the index $n$ which the state process is assigned by the mapping $\mathcal{I}$:

$$\nu^*(\omega \overset{\theta}{\oplus} \omega', t) := \begin{cases} \nu^\eta(\omega, t), & \text{if } t \in [0, \theta(\omega)], \\ \nu^{\mathcal{N}(\omega)}(\omega', t - \theta(\omega)), & \text{if } t > \theta(\omega), \end{cases}$$

with $\mathcal{N}(\omega) = \mathcal{I}((X_{\theta(\omega)}, Y_{\theta(\omega)})^{\nu^\eta, x, y})$. This construction ensures that we have $\nu^* \in \Theta^\lambda(x, y)$. Hence, it follows from the definitions of the value function and $\nu^*$, (3.6) and $v^\lambda \geq \varphi$ (which holds by definition of $\varphi$), as well as (3.5) and (3.3) that

$$\begin{aligned} v^\lambda(x, y) &\geq \mathbb{E}\left[\int_0^\infty e^{-\beta t} U\left(c_t^{\nu^*}\right) dt\right] \\ &= \mathbb{E}\left[\int_0^\theta e^{-\beta t} U\left(c_t^\eta\right) dt + e^{-\beta \theta} \int_0^\infty e^{-\beta t} U\left(c_t^\mathcal{N}\right) dt\right] \\ &\geq \mathbb{E}\left[\int_0^\theta e^{-\beta t} U(c_t^\eta) dt + e^{-\beta \theta}(\varphi(\zeta((X_\theta, Y_\theta)^{\nu^\eta, x, y})) - \eta)\right] \\ &\geq \mathbb{E}\left[\int_0^\theta e^{-\beta t} U(c_t^\eta) dt + e^{-\beta \theta}(\varphi((X_\theta, Y_\theta)^{\nu^\eta, x, y}) - 2\eta)\right] \\ &\geq \mathbb{V} - 3\eta. \end{aligned}$$

Since $\eta$ was arbitrary this establishes (3.2), thereby completing the proof. □

## 3.2  $v^\lambda$ is a viscosity solution of (2.4)

We first state and prove some facts about the intervention operator $\mathbf{M}$ from (2.5), which are needed in the subsequent proofs. Throughout, $\psi_*$ and $\psi^*$ will denote the lower and upper-semicontinuous envelopes of a locally bounded function $\psi$, respectively.

**Lemma 3.1.** *Suppose $\lambda_f, \lambda_p > 0$. Let $\varphi : K_\lambda \to \mathbb{R}$. Then*

  (i) *If $\varphi$ is upper semi-continuous, then $\mathbf{M}\varphi$ is upper semi-continuous.*

  (ii) *If $\varphi$ is continuous, then $\mathbf{M}\varphi$ is continuous.*

*Proof.* The proof can be found in [37]  □

**Remark 3.2.** *When $\lambda_p = 0$, the above lemma is no longer true. To see where the argument breaks down consider the smooth function $h$ on $\mathbb{R}_+^2$ defined by*

$$h(x,y) = \begin{cases} h_0(x-y), & x+y > 2 \\ h_{\tan(\pi(2-x-y))}(x-y), & 1 < x+y \leq 2 \\ 0, & x+y \leq 1. \end{cases}$$

*where $h_\zeta : \mathbb{R} \to [0,1]$ is the standard smooth bump function centered at $\zeta$, i.e. with peak $h_\zeta(\zeta) = 1$. Suppose that $\lambda_f = 1$. Consider the sequence $\zeta_n := (1 + \frac{1}{n}, 1 + \frac{1}{n}) \to (1,1)$. Then, $\limsup_{n\to\infty} \mathbf{M}h(\zeta_n) = 1 > 0 = \mathbf{M}h(1,1)$ which demonstrates that upper-semicontinuity is not preserved by $\mathbf{M}$. Compactness of iso-wealth lines, however, would preclude us from pushing bumps out to infinity.*

The following lemma is needed in the case when $\lambda_f > 0$ and $\lambda_p = 0$. A fundamental difficulty arises in the pure fixed cost case as the set of attainable portfolios at a fixed wealth level is no longer compact as it was in the case when $\lambda_f > 0$ and $\lambda_p > 0$.

**Lemma 3.3.** *Suppose $\varphi : K_\lambda \to \mathbb{R}$ satisfies $\sup_{z \in K} \|\varphi(z, \cdot)\|_\infty < \infty$ for any non-empty compact set $K \subset \mathbb{R}_+$.*

  (i) *If $\varphi$ is lower semi-continuous, then $\mathbf{M}\varphi$ is lower semi-continuous. In particular, if $\varphi \geq \mathbf{M}\varphi$, then $\varphi_* \geq \mathbf{M}\varphi_*$.*

  (ii) *Let $\varphi \in C^1(K_\lambda)$. If $(z,\xi) \mapsto D_\xi \varphi(z,\xi)$ is compactly supported on $C \times \mathbb{R}^d$ for any compact set $C \subset R_+$, then $\mathbf{M}\varphi$ is upper semi-continuous.*

*Proof.* See [1] for the proof.  □

The following definition in an adaptation of the one given in [45, 46] to the current problem. The main difference between the classical viscosity solution and the one below is that for a constrained solution the sub-solution property extends to the closed domain $\mathfrak{D}$.

**Definition 3.4.** *We say that $u$ is a viscosity subsolution on $\mathfrak{D}$ if for each $\zeta_0 \in \mathfrak{D}$ and for every upper-semicontinuous function $\varphi$ such that $\varphi$ is locally $C^2$ at $\zeta_0$ and $0 = (u^* - \varphi)(\zeta_0) = \max_{\zeta \in \mathfrak{D}}(u^* - \varphi)(\zeta)$ there holds*

$$\min\left\{\beta\varphi(\zeta_0) - \tilde{U}(\varphi_x(\zeta_0)) - \mathscr{L}\varphi(\zeta_0), (\varphi - \mathbf{M}\varphi)_*(\zeta_0)\right\} \leq 0.$$

*We say that $u$ is a viscosity supersolution on $\mathring{\mathfrak{D}}$ if for each $\zeta_0 \in \mathring{\mathfrak{D}}$ and every smooth $\varphi$ such that $0 = (u_* - \varphi)(\zeta_0) = \min_{\zeta \in \mathring{\mathfrak{D}}}(u_* - \varphi)(\zeta)$ there holds*

$$\min\left\{\beta\varphi(\zeta_0) - \tilde{U}(\varphi_x(\zeta_0)) - \mathscr{L}\varphi(\zeta_0), (\varphi - \mathbf{M}\varphi)(\zeta_0)\right\} \geq 0.$$

*We say that $u$ is a constrained viscosity solution on $\mathfrak{D}$, if it is a subsolution on $\mathfrak{D}$ and a supersolution in $\mathring{\mathfrak{D}}$.*

**Remark 3.5.** *In the given definition of viscosity subsolutions we have chosen a relaxation of the conditions on the test functions, letting them be merely upper-semicontinuous outside of some neighborhood. This is needed in the proof of Theorem 4.3, due to the global behavior of the operator $\mathbf{M}$.*

**Remark 3.6.** *When both $\lambda_f, \lambda_p > 0$, then the equations are continuous, i.e. the lower envelope of the equation for viscosity subsolutions is not needed. The reason is that the operator $\mathbf{M}$ preserves upper-semicontinuity, in other words, the quantity in the equation is already lower semicontinuous.*

We are now ready to tackle the proof of Theorem 2.3, which we split into two lemmata:

**Lemma 3.7.** *The value function $v^\lambda$ is a viscosity supersolution of the Dynamic Programming Equation (2.4) on $\mathring{\mathfrak{D}}$.*

*Proof.* Let $(x_0, y_0) \in \mathring{\mathfrak{D}}$. and let $\varphi$ be a smooth function on $\mathcal{O}_\lambda := \mathcal{O}_\lambda(x_0, y_0; 2(x_0 + y_0 \cdot \mathbf{1}_d))$ satisfying

$$0 = (v_*^\lambda - \varphi)(x_0, y_0) = \min\{(v_*^\lambda - \varphi)(x', y') : (x', y') \in \mathcal{O}_\lambda\}.$$

Using Lemma 3.3 and the inequality $v_*^\lambda \geq \varphi$ on $\mathcal{O}_\lambda$ we obtain

$$\varphi(x_0, y_0) = v_*^\lambda(x_0, y_0) \geq \mathbf{M}v_*^\lambda(x_0, y_0) \geq \mathbf{M}\varphi(x_0, y_0).$$

Therefore, it remains to show that

$$\left(\beta\varphi - \widetilde{U}(\varphi_x) - \mathscr{L}\varphi\right)(x_0, y_0) \geq 0.$$

Assume to the contrary that

$$\left(\beta\varphi - U(c^*) + c^*\varphi_x - \mathscr{L}\varphi\right)(x_0, y_0) < 0,$$

for some $c^* > 0$, and set $\phi(x, y) := \varphi(x, y) - \epsilon(|x - x_0|^4 + \|y - y_0\|^4)$. Then, for $\epsilon > 0$ and $r > 0$ small enough, continuity yields

$$\big(\beta\phi - U(c^*) + c^*\phi_x - \mathscr{L}\phi\big)(x, y) < 0, \quad \forall (x, y) \in \overline{B}_r(x_0, y_0) \subset \mathcal{O}_\lambda.$$

Select a convergent sequence of points $(x_n, y_n, v^\lambda(x_n, y_n)) \to (x_0, y_0, v^\lambda_*(x_0, y_0))$ and denote by $(X^n_t, Y^n_t) := (X_t, Y_t)^{x_n, y_n}$ the portfolio process starting at $(x_n, y_n)$ under the consumption-only strategy $c_t \equiv c^*$. Define

$$H^n := \inf\{t \geq 0 : (X^n_t, Y^n_t) \notin \overline{B}_r(x_0, y_0)\}$$

and note that $\liminf_{n\to\infty} \mathbb{E}[H^n] > 0$. Hence, there exists $\delta > 0$ such that $\mathbb{E}[e^{-\beta H^n}] > \delta$, for all $n$ sufficiently large. Itô's formula gives

$$\phi(x_n, y_n)$$
$$= \mathbb{E}\left[e^{-\beta H^n}\phi(X^n_{H^n}, Y^n_{H^n}) + \int_0^{H^n} e^{-\beta s}(\beta\phi + c^*\phi_x - \mathscr{L}\phi)(X^n_s, Y^n_s)ds\right]$$
$$\leq \mathbb{E}\left[e^{-\beta H^n}\phi(X^n_{H^n}, Y^n_{H^n}) + \int_0^{H^n} e^{-\beta s}U(c^*)ds\right].$$

By construction of $\phi$, there exists $\eta > 0$ such that we have $\varphi \geq \phi + \eta$ on $\mathcal{O}_\lambda \backslash \overline{B}_r(x_0, y_0)$. Hence:

$$\phi(x_n, y_n) \leq \mathbb{E}\left[e^{-\beta H^n}\varphi(X^n_{H^n}, Y^n_{H^n}) + \int_0^{H^n} e^{-\beta s}U(c^*)ds\right] - \delta\eta.$$

Taking into account $(v^\lambda - \phi)(x_n, y_n) \to 0$, we note that for $n$ large enough

$$v^\lambda(x_n, y_n) \leq \mathbb{E}\left[e^{-\beta H^n}\varphi(X^n_{H^n}, Y^n_{H^n}) + \int_0^{H^n} e^{-\beta s}U(c^*)ds\right] - \frac{\delta\eta}{2}.$$

This contradicts the weak dynamic programming principle (3.2) for $v^\lambda$, thereby completing the proof. □

The image of an arbitrary smooth function under $\mathbf{M}$ is upper semicontinuous only under additional assumptions (compare 3.3(ii)). As is customary in the theory of viscosity solutions (cf., e.g., section 9 of [10]), the viscosity subsolution property in the following lemma is therefore formulated in terms of the lower semicontinuous envelope of the DPE:

**Lemma 3.8.** *The value function $v^\lambda$ is a viscosity subsolution of*

$$\min\left\{\beta v^\lambda - \widetilde{U}(v^\lambda_x) - \mathscr{L}v^\lambda, (v^\lambda - \mathbf{M}v^\lambda)_*\right\} = 0 \quad \text{on } \mathfrak{D}.$$

*Proof. Step 1.* Throughout the proof, $C > 0$ denotes a generic constant that may vary from line to line. We argue by contradiction. Let $(x_0, y_0) \in \mathfrak{D}$ and let $\varphi$ be an upper-semicontinuous and bounded function on $\mathcal{O}_\lambda(x_0, y_0; 2(x_0 + y_0 \cdot \mathbf{1}_d))$ which is locally $C^2$ at $(x_0, y_0)$ and satisfies

$$0 = ((v^\lambda)^* - \varphi)(x_0, y_0) = \max\{((v^\lambda)^* - \varphi)(x', y') : (x', y') \in \mathcal{O}_\lambda\}.$$

Suppose that for some $\eta > 0$, we have

$$\min\{\beta\varphi - \mathscr{L}\varphi - \tilde{U}(\varphi_x), (\varphi - \mathbf{M}\varphi)_*\}(x_0, y_0) > \eta.$$

By lower semi-continuity, there is a small rectangular neighborhood

$$N = N(x_0, y_0, \rho) := \left\{(x, y) \in \mathcal{O}_\lambda : \max_{i=1,\ldots,d}\{|x - x_0|, |y^i - y_0^i|\} < \rho\right\}$$

on which $\varphi$ is $C^2$ and satisfies

$$\min\{\beta\varphi - \mathscr{L}\varphi + c\varphi_x - U(c), \varphi - \mathbf{M}\varphi\}(x, y) > \eta, \qquad (3.7)$$

for all $c > 0$, $(x, y) \in N$.

*Step 2.* Choose a sequence $N \ni (x_n, y_n) \to (x_0, y_0)$ for which $v^\lambda(x_n, y_n)$ converges to $(v^\lambda)^*(x_0, y_0)$. At each of these points choose a $\frac{1}{n}$-optimal control $\nu^n \in \Theta^\lambda(x_n, y_n)$. We denote by $c_t^n$ and $\tau^n$ the consumption process and the first impulse time of $\nu^n$, respectively, and write $(X_t^n, Y_t^n) := (X_t, Y_t)^{\nu^n, x_n, y_n}$ for the corresponding controlled process. Denote by $(\Xi_t^n) \in \mathbb{R}^2$ the same process, but without trading, i.e., the process starting at $(x_n, y_n)$ and with consumption $c^n$.

Define the stopping times

$$H^n := \inf\{t \geq 0 : \Xi_t^n \notin N\} \wedge 1,$$

and

$$\theta^n := H^n \wedge \tau^n.$$

We can further decompose $H^n = \underline{H}^n \wedge \overline{H}^n \wedge 1$, where

$$\underline{H}^n := \inf\{t \geq 0 : \Xi_t^n \in \partial N \cap \{x_0 - \rho\} \times \mathbb{R}^d\},$$

and

$$\overline{H}^n := \inf\{t \geq 0 : \Xi_t^n \in \partial N \cap \{x : x > x_0 - \rho\} \times \mathbb{R}^d\}.$$

The stopping time $\overline{H}^n$ captures exit by diffusion and $\underline{H}^n$ represents exit by consumption.

*Step 3.* Write

$$h(c, x, y) := I(c, x, y) - \sup_{\hat{c} > 0} I(\hat{c}, x, y),$$

where

$$I(c, x, y) := -\beta\varphi(x, y) + \mathscr{L}\varphi(x, y) - c\varphi_x(x, y) + U(c).$$

Note that $I(c, x, y) < 0$ for all $c \in \mathbb{R}_+$ and $(x, y) \in N$ by (3.7). Setting $c^*(x, y) = (U')^{-1}(\varphi_x(x, y))$, it follows that

$$h(c, x, y) = I(c, x, y) - I\big(c^*(x, y), x, y\big) \leq 0.$$

By smoothness of $\varphi$ and $c^*$ and compactness of $N$, there exists $L_\rho > 0$ with $|I(c^*(x,y), x, y)| \leq L_\rho$, for all $(x,y) \in N$. On the other hand, there is $\alpha > 0$ such that $I(c, x, y) \leq -\alpha c$, for all $c > 0$. This leads to the upper bound

$$h(c, x, y) \leq (-\alpha c + L_\rho) \wedge 0, \qquad \text{for all } c > 0, (x,y) \in N. \tag{3.8}$$

As we only consider times $t$ up to $\theta^n$, we can assume without loss of generality that $c_t^n = c^*(X_t^n, Y_t^n)$ for $t \in (\theta^n, H^n]$. Together with (3.8) we obtain

$$\begin{aligned}
\mathbb{E}\left[\int_0^{\theta^n} -e^{-\beta t} h(c_t^n, X_t, Y_t) dt\right] &= \mathbb{E}\left[\int_0^{H^n} -e^{-\beta t} h(c_t^n, X_t, Y_t) dt\right] \\
&\geq C\alpha \mathbb{E}\left[\int_0^{H^n} e^{-rt} c_t^n dt\right] - L_\rho \mathbb{E}[H^n]. \\
&\geq C\alpha \mathbb{E}\left[\int_0^{\theta^n} e^{-rt} c_t^n \mathbf{1}_{\{\theta^n = \underline{H}^n\}} dt\right] \\
&\quad - L_\rho \mathbb{E}[H^n],
\end{aligned} \tag{3.9}$$

where the first inequality uses (3.8) and changes the discount factor.

*Step 4.* Set $\zeta_t^n := (X_t^n, Y_t^n)$. Weak dynamic programming (3.1) implies

$$\begin{aligned}
v^\lambda(x_n, y_n) &\leq \frac{1}{n} + \mathbb{E}\left[\int_0^{\theta^n} e^{-\beta t} U(c_t^n) dt + e^{-\beta \theta^n} \varphi(\zeta_{\theta^n}^n)\right] \\
&\leq \frac{1}{n} + \varphi(x_n, y_n) + \mathbb{E}\left[\int_0^{\theta^n} e^{-\beta t} I(c_t^n, \zeta_t^n) dt\right] \\
&\quad + \mathbb{E}\left[e^{-\beta \theta^n}(\varphi(\zeta_{\theta^n}^n) - \varphi(\zeta_{\theta^n-}^n))\mathbf{1}_{\{\theta^n = \tau^n\}}\right] \\
&\leq \frac{1}{n} + \varphi(x_n, y_n) + \mathbb{E}\left[\int_0^{\theta^n} e^{-\beta t} I(c_t^*(\zeta_t^n), \zeta_t^n) dt\right] \\
&\quad + \mathbb{E}\left[\int_0^{\theta^n} e^{-\beta t} h(c_t^n, \zeta_t^n) dt\right] - C\eta \mathbb{P}[\theta^n = \tau^n] \\
&\leq \frac{1}{n} + \varphi(x_n, y_n) - CL_\rho \eta \mathbb{E}[\theta^n] - C\eta \mathbb{P}[\theta^n = \tau^n] \\
&\quad + \mathbb{E}\left[\int_0^{\theta^n} e^{-\beta t} h(c_t^n, \zeta_t^n) dt\right].
\end{aligned}$$

Since $v^\lambda(x_n, y_n) - \varphi(x_n, y_n) - \frac{1}{n} \to 0$ as $n \to \infty$ and since the other terms on the right-hand side are negative, they must each vanish as $n$ tends to infinity.

*Step 5.* We derive a contradiction using that

$$\lim_{n \to \infty} \max\left\{\mathbb{E}[\theta^n], \, \mathbb{P}[\theta^n = \tau^n], \, \mathbb{E}\left[\int_0^{\theta^n} -e^{-\beta t} h(c_t^n, \zeta_t^n) dt\right]\right\} = 0. \tag{3.10}$$

It will be useful to have $\mathbb{E}[H^n] \to 0$ and $\mathbb{P}[\theta^n = \underline{H}^n] \to 1$. To prove these statements, first observe that

$$1 = \mathbb{P}[\theta^n = H^n] + \mathbb{P}[\theta^n = \tau^n] - \mathbb{E}[\mathbf{1}_{\{\theta^n = H^n\}} \mathbf{1}_{\{\theta^n = \tau^n\}}]$$

where $\mathbb{E}[\mathbf{1}_{\{\theta^n = H^n\}} \mathbf{1}_{\{\theta^n = \tau^n\}}] \to 0$. This implies that $\mathbb{P}[\theta^n = H^n] \to 1$. In addition,

$$\mathbb{E}[\theta^n] = \int_{\{\theta^n = H^n\}} H^n d\mathbb{P} + \int_{\{\theta^n \neq H^n\}} \theta^n d\mathbb{P},$$

where $\int_{\{\theta^n = H^n\}} \theta^n d\mathbb{P} \to 0$. This implies

$$\mathbb{E}[H^n] - \int_{\{\theta^n \neq H^n\}} H^n d\mathbb{P} \to 0.$$

Since $|H^n| \leq 1$ by definition and since $\mathbb{P}[\theta^n \neq H^n] \to 0$ according to the observations above, we obtain $\mathbb{E}[H^n] \to 0$. The statement that $\mathbb{P}[\theta^n = \underline{H}^n] \to 1$ follows the fact that $H^n = \overline{H}^n \wedge \underline{H}^n \wedge 1 \to 0$ and $\overline{H}^n$ is the exit time from $N$ of a diffusion process started at $(x_n, y_n) \to (x_0, y_0)$.

*Step 6.* As a consequence, if $(x_0, y_0)$ is an interior point of $\mathfrak{D}$ and $\rho$ chosen small enough,

$$\mathbb{E}\left[\int_0^{\theta^n} e^{-rt} c_t^n \mathbf{1}_{\{\theta^n = \underline{H}^n\}} dt\right] \to \rho,$$

which follows from the simple observation that, for any fixed $n$, the term inside the expectation represents the amount of discounted consumption needed for cash in the bank account to decrease from $x_n$ to $x_0 - \rho$.

However by (3.10) and (3.9), we must have

$$0 = \lim_{n \to \infty} \mathbb{E}\left[\int_0^{\theta^n} -e^{-\beta t} h(c_t, X_t, Y_t) dt\right] \geq C\alpha\rho - L_\rho \lim_{n \to \infty} \mathbb{E}[H^n] = C\alpha\rho > 0,$$

which is a contradiction.

*Step 7.* In the case $U(0) > -\infty$, we need to discuss the solution at the boundary. Note that if $(x_0, y_0) \in \partial K_\lambda \setminus S = K_\lambda \cap \mathbb{R}^{d+1}_{\geq 0}$ (recall (3.4)), then it is immediate from dynamic programming that

$$\beta\varphi(x_0, y_0) - \tilde{U}(\varphi_x)(x_0, y_0) - \mathscr{L}\varphi(x_0, y_0) = 0.$$

Suppose $(x_0, y_0) \in S$. We require that $\varphi$ is taken so that Lemma 3.3 (ii) applies. Note that the previous steps of the proof are agnostic to this requirement. If $(v^\lambda)^*(x_0, y_0) = v^\lambda(x_0, y_0)$, then

$$(\varphi - \mathbf{M}\varphi)(x_0, y_0) \leq \varphi(x_0, y_0) - \mathbf{M}v^\lambda(x_0, y_0)$$
$$= v^\lambda(x_0, y_0) - v^\lambda(\mathbf{0}_{d+1}) = 0,$$

since the only admissible control at $(x_0, y_0)$ is liquidation.

On the other hand, if $(v^\lambda)^*(x_0, y_0) \neq v^\lambda(x_0, y_0)$, then suppose by way of contradiction that

$$\min\{\beta\varphi - \mathscr{L}\varphi - \tilde{U}(\varphi_x),\ (\varphi - \mathbf{M}\varphi)_*\}(x_0, y_0) > \eta,$$

for some $\eta > 0$. Repeat Steps 1–5 with a sufficiently small $\rho$ and the stopping times $\underline{H}^n$ and $\overline{H}^n$ re-defined to be

$$\underline{H}^n = \inf\{t \geq 0 : \Xi_t^n \in \partial N \cap S\}$$

and

$$\overline{H}^n = \inf\{t \geq 0 : \Xi_t^n \in \partial N \setminus S\}.$$

Repeating these steps is possible since $v^\lambda$ is non-decreasing in each argument, and the sequence in Step 2 can therefore be contained in the interior of $\mathfrak{D}$. Note that at time $\underline{H}^n$, the process is in $S$, where the only admissible control is to trade to the origin. Hence, $\tau^n \leq \underline{H}^n$, and thus also $\tau^n = \theta^n$ on the set $\{\theta^n = \underline{H}^n\}$. By Step 5, this implies

$$0 = \lim_{n \to \infty} \mathbb{P}[\theta^n = \tau^n] = \lim_{n \to \infty} \mathbb{P}[\theta^n = \underline{H}^n] = 1,$$

which is clearly a contradiction. Thus,

$$\min\{\beta\varphi - \mathscr{L}\varphi - \tilde{U}(\varphi_x),\ (\varphi - \mathbf{M}\varphi)_*\}(x_0, y_0) \leq 0,$$

which is the desired inequality. □

## 4 Comparison

In this section, we assume that $\lambda_f > 0$ and $\lambda_p \geq 0$ and we are now considering more general utility functions.

**Proposition 4.1.** *Let $U$ be an increasing, smooth, strictly concave utility function on $\mathbb{R}_+$. Then its Legendre-Fenchel transform, defined by*

$$\tilde{U}(\tilde{c}) := \sup_{c>0}\{U(c) - c\tilde{c}\}, \quad \tilde{c} \in \mathbb{R},$$

*is decreasing.*

*Proof.* A simple calculus argument shows that

$$\frac{d}{d\tilde{c}}\tilde{U}(\tilde{c}) = -(U')^{-1}(\tilde{c}) < 0,$$

where the negativity follows by the assumptions on $U$. □

We now aim to reformulate the definition of the intervention operator $\mathbf{M}$ so as to algebraically manipulate the strategies. For each $\zeta \in \mathfrak{D}$, we define the sets

$$\mathcal{I}(\zeta) := \{\nu = (m^1 - \sum_{i=2}^{d+1}\lambda_p|m^i| - \lambda_f, m^2, \ldots, m^{d+1}) : \sum_{i=1}^{d+1} m^i = 0, \zeta + \nu \in \mathfrak{D}\}.$$

Then, the intervention operator satisfies

$$\mathbf{M}\psi(\zeta) = \sup_{\nu \in \mathcal{I}(\zeta)} \{\psi(\zeta + \nu)\}.$$

By convention, if $\mathcal{I}(\zeta) = \emptyset$, then $\mathbf{M}\psi(\zeta) = -\infty$.

**Lemma 4.2.** *If $\zeta \leq \hat{\zeta}$, then $\mathcal{I}(\zeta) \subset \mathcal{I}(\hat{\zeta})$.*

*Proof.* If $\mathcal{I}(\zeta) = \emptyset$, then the assertion is trivially true. Otherwise, if there exists $\nu \in \mathcal{I}(\zeta)$. Since $0 \leq (\zeta + \nu) \cdot \mathbf{1}_{d+1} < (\hat{\zeta} + \nu) \cdot \mathbf{1}_{d+1}$, it follows that $\hat{\zeta} + \nu \in \mathfrak{D}$. Hence $\nu \in \mathcal{I}(\hat{\zeta})$  $\square$

**Theorem 4.3.** *Let $u$ be an upper-semicontinuous sub-solution on $\mathfrak{D}$, and $v$ be a lower-semicontinuous super-solution in $\mathring{\mathfrak{D}}$ and set $\overline{\alpha} := (\alpha, 0, \ldots, 0) \in \mathbb{R}^{d+1}$ for some $\alpha > 0$. Suppose that $u$ and $v$ are non-decreasing in the variables $(x, y^1, \ldots, y^d)$. If*

$$\inf_{\eta \in \mathfrak{D} \setminus \mathring{\mathfrak{D}}} v(\eta + \overline{\alpha}) > -\infty \qquad (4.1)$$

*and if there exists some $\hat{\gamma} \leq \gamma$ large enough to also satisfy Assumption 2.1 as well as a $C > 0$ such that*

$$u(x, y) \leq C(1 + |x + y \cdot \mathbf{1}_d|^{1-\hat{\gamma}}), \quad \forall (x, y) \in \mathring{\mathfrak{D}}, \qquad (4.2)$$

*then $u(\eta) \leq v(\eta + \overline{\alpha})$, for all $\eta \in \mathfrak{D}$.*

Before presenting the proof, let us state the theorem describing the relevance to $v^\lambda$ and how it relates to comparison:

**Theorem 4.4.** *Let $u$ and $v$ be two constrained viscosity solutions, both satisfying (4.1) and (4.2) for all $\alpha > 0$. Then*

*(i) $u^*(\eta) = v^*(\eta)$ for all $\eta \in \mathfrak{D}$.*

*(ii) $u_*(\eta) = v_*(\eta)$ for all $\eta \in \mathring{\mathfrak{D}}$.*

*In particular, the value function $v^\lambda$ satisfies these conditions whenever the asymptotic elasticity of the utility function $U$ is smaller than 1.*

*Proof.* We begin by proving the following:

$$u^*(\eta) \leq (v_*)^*(\eta) \text{ and } v^*(\eta) \leq (u_*)^*(\eta) \quad \forall \eta \in \mathfrak{D} \qquad (4.3)$$

as well as

$$(u^*)_*(\eta) \leq v_*(\eta) \text{ and } (v^*)_*(\eta) \leq u_*(\eta) \quad \forall \eta \in \mathring{\mathfrak{D}}. \qquad (4.4)$$

Theorem 4.3 holds for $u^*$ and $v_*$ as $u$ and $v$, respectively. Thus,

$$u^*(\eta) \leq \limsup_{\alpha \to 0} v_*(\eta + \overline{\alpha}) \leq (v_*)^*(\eta),$$

for any choice of $\eta \in \mathfrak{D}$. Similarly, for $\eta \in \mathring{\mathfrak{D}}$,

$$v_*(\eta) \geq \liminf_{\alpha \to 0} u^*(\eta - \overline{\alpha}) \geq (u^*)_*(\eta).$$

Interchanging $u$ and $v$ above, we obtain the other halves of the statements.

Employing (4.3) twice yields

$$u^* \leq (v_*)^* \leq v^* \leq (u_*)^* \leq u^*,$$

implying $u^* = v^*$ on $\mathfrak{D}$. Moreover, by (4.4),

$$v_* \geq (u^*)_* \geq u_* \geq (v^*)_* \geq v_*,$$

implying $v_* = u_*$ in $\mathring{\mathfrak{D}}$.

Finally, by Assumption 2.1 and the following Remark 2.2, condition (4.2) is satisfied for $v^\lambda$. Now choose any $\alpha > 0$. Then, for any $\eta \in \mathfrak{D} \subset K_\lambda$,

$$v^\lambda(\eta + \overline{\alpha}) \geq v^\lambda(\overline{\alpha}) \geq \frac{U(r\alpha)}{\beta} > -\infty,$$

which follows from the definition of $K_\lambda$ and by choosing the consumption to be the interest $r\alpha$. Hence, (4.1) is also satisfied. $\square$

Note that if $u$ is a viscosity solution satisfying (4.1) and (4.2), but not necessarily for all $\alpha > 0$, the same type of argument as in the proof yields

$$u^*(\eta) \leq (v^\lambda)^*(\eta), \quad \forall \eta \in \mathfrak{D}$$

and

$$u_*(\eta) \leq (v^\lambda)_*(\eta), \quad \forall \eta \in \mathring{\mathfrak{D}}.$$

This theorem also justifies the slightly unorthodox statement of comparison in Theorem 4.3. Indeed, a traditional comparison formulation would always imply that the constrained viscosity solutions are unique and therefore continuous. Hence, $v^\lambda$ is truly discontinuous, so a traditional comparison formulation cannot be expected. Therefore, the above uniqueness result is the best one could prove, as the viscosity solutions do not distinguish the semi-continuous envelopes.

*Proof of Theorem 4.3. Step 1.* We just prove the comparison theorem for the case of pure fixed costs ($\lambda_f > 0, \lambda_p = 0$). In the case when both $\lambda_f, \lambda_p > 0$, the solvency region is compact for all fixed wealth levels $z$. This compactness is lost when $\lambda_p = 0$ and the proof in the case of pure fixed costs is in fact more complicated (and can easily be carried over to the more general transaction costs case, *mutatis mutandis*). We will abuse notation and write $\lambda := \lambda_f$.

*Step 2.* Suppose by way of contradiction that $u(\eta_0) - v(\eta_0 + \overline{\alpha}) > 0$ at some point $\eta_0 \in \mathfrak{D}$. Note that if $\eta_0 \in \partial \mathfrak{D}$, then $u(\eta_0) > -\infty$. Take any $\gamma' < \min\{\hat{\gamma}, 1\}$ and choose $\epsilon > 0$ small enough so that $u(\eta_0) - v(\eta_0 + \overline{\alpha}) - \epsilon(z(\eta_0 + \overline{\alpha}))^{1-\gamma'} > 0$, where $z(\eta) := \eta \cdot \mathbf{1}_{d+1}$. Set

$$\Psi^\iota(\eta, \xi; \gamma') := u(\eta) - v(\xi + \overline{\alpha}) - \epsilon(z(\xi + \overline{\alpha}))^{1-\gamma'} - \frac{1}{2}\iota|\eta - \xi|^2.$$

Observe that due to the growth restrictions (4.2) on $u$ and the finiteness assumption on $v$, we have $\Psi^\iota(\eta, \xi; \gamma') \to -\infty$ uniformly in $\eta$ as $|\xi| \to \infty$. As a result, there exists $L := L(\gamma') > 0$ such that whenever $\hat{\xi} \in \mathfrak{D}$ satisfies $z(\hat{\xi} + \overline{\alpha}) > L$, then $\Psi^\iota(\eta, \hat{\xi}; \gamma') < 0$ for all $\eta \in \mathfrak{D}$. This construction compactifies the problem in the wealth direction, however maximizers of $\Psi^\iota(\eta, \xi; \gamma')$ may not exist since the strip $\{\eta' \in \mathfrak{D} : z(\eta' + \overline{\alpha}) \leq L\}$ is not compact in the case when there are no proportional costs (i.e. $\lambda_p = 0$).

*Step 3.* The conditions on $u$ and $v$ imply that $\sup_{\eta, \xi \in \mathfrak{D}} \Psi^\iota(\eta, \xi; \epsilon, \gamma') < \infty$. Consequently, there exists a maximizing sequence of points $(\eta_j, \xi_j) \in \mathfrak{D} \times \mathfrak{D}$ for which $\Psi^\iota(\eta_j, \xi_j; \gamma') \to \sup_{\eta, \xi \in \mathfrak{D}} \Psi^\iota(\eta, \xi; \gamma')$ as $j \to \infty$. Let us introduce a smooth bump function $h_j : \mathfrak{D} \to \mathbb{R}$ such that its peak satisfies $h_j(\xi_j + \overline{\alpha}) = 1$ and its support is contained in a relatively open ball in $\mathfrak{D}$ of radius $r < \frac{1}{2}\min\{\alpha, \lambda\}$. The bump function $h_j$ then satisfies a simple $C^2$ estimate

$$|\beta h - \mathscr{L} h| \leq c\|h_j\|_{C^2(\mathfrak{D})},$$

for some generic constant $c > 0$ depending on the market parameters and $r$. Define

$$\delta_j := \frac{3}{2}\left(\sup_{\eta, \xi \in \mathfrak{D}} \Psi^\iota(\eta, \xi; \gamma') - \Psi^\iota(\eta_j, \xi_j; \gamma')\right).$$

Clearly, $\delta_j > 0$ and $\delta_j \to 0$ as $j \to \infty$. Finally, let us observe that

$$\Psi^{\iota, \delta_j}(\eta, \xi; \gamma') := \Psi^\iota(\eta, \xi; \gamma') + \delta_j h_j(\xi + \overline{\alpha})$$

has a maximizer, say $(\hat{\eta}_j, \hat{\xi}_j)$, which is near $(\eta_j, \xi_j)$. We will forego the additional notation and just assume the maximum is at $(\eta_j, \xi_j)$.

*Step 4.* We now construct a strict super-solution on the domain $\mathfrak{D}_{\alpha, L(\gamma')} := \{\xi \in \mathfrak{D} : \alpha \leq z(\xi) < L(\gamma')\}$. We claim that provided $0 < \hat{\gamma} - \gamma'$ is sufficiently small, then for all $\epsilon > 0$ and $j$ sufficiently large, the function $v^{\epsilon, \delta_j}(\xi) := v(\xi) + \epsilon(z(\xi))^{1-\gamma'} - \delta_j h_j(\xi)$ is a strict super-solution of the dynamic programming equation on $\mathfrak{D}_{\alpha, L(\gamma')} \cap \mathring{\mathfrak{D}}$. Let $\varphi$ be a smooth test function touching $v$ from below at some point $\hat{\xi}$. Let us write $g(\xi) := (z(\xi))^{1-\gamma'}$. We begin by stating a few facts about $g$. First, note that by the finiteness criterion for the Merton value function, given in Assumption 2.1, for $\gamma'$ sufficiently close to $\min\{\hat{\gamma}, 1\}$ there exists $\rho_1 := \rho_1(\alpha, L(\gamma')) > 0$ such that

$$\beta g(\xi) - \mathscr{L} g(\xi) > \rho_1, \quad \forall \xi \in \mathfrak{D}_{\alpha, L(\gamma')}.$$

We fix such a $\gamma'$ for the rest of the proof and will shortly suppress it from the notation. Next, it is also clear that

$$\partial_x(\epsilon g(\xi) - \delta_j h_j(\xi)) > 0, \text{ holds for all } \xi \in \mathfrak{D}_{\alpha, L(\gamma')}$$

provided $j$ is sufficiently large (i.e. $\delta_j$ sufficiently small). Then

$$\beta(\varphi + \epsilon g - \delta_j h_j) - \mathscr{L}(\varphi + \epsilon g - \delta_j h_j) - \tilde{U}(\varphi_x + \partial_x(\epsilon g - \delta_j h_j))$$
$$= \beta\varphi - \mathscr{L}\varphi - \tilde{U}(\varphi_x) + \epsilon(\beta g - \mathscr{L}g) - \delta_j(\beta h_j - \mathscr{L}h_j)$$
$$\quad + \tilde{U}(\varphi_x) - \tilde{U}(\varphi_x + \partial_x(\epsilon g - \delta_j h_j))$$
$$\geq \epsilon(\beta g - \mathscr{L}g) - \delta_j c \|h\|_{C^2}$$
$$\geq \frac{\epsilon \rho_1}{2}, \quad \text{for } j \text{ sufficiently large.}$$

Finally, we show that there exists $\rho_2 := \rho_2(\alpha, L(\gamma')) > 0$ such that $v^{\epsilon,\delta_j} - \mathbf{M}v^{\epsilon,\delta_j} \geq \frac{1}{2}\epsilon\rho_2$ on $\mathfrak{D}_{\alpha,L(\gamma')}$. Indeed,

$$v^{\epsilon,\delta_j} - \mathbf{M}v^{\epsilon,\delta_j} = v + \epsilon g - \delta_j h_j - \mathbf{M}(v + \epsilon g - \delta_j h_j)$$
$$\geq v - \mathbf{M}v + \epsilon(g - \mathbf{M}g) - \delta_j h_j - \mathbf{M}(-\delta_j h_j)$$
$$\geq \epsilon \inf_{\xi \in \mathfrak{D}_{\alpha,L(\gamma')}} (g - \mathbf{M}g)(\xi) - \delta_j h_j$$
$$\geq \epsilon \rho_2 - \delta_j c \|h_j\|_{C^2}$$
$$\geq \frac{1}{2}\epsilon \rho_2, \quad \text{for } j \text{ sufficiently large.}$$

Setting $\rho := \min\{\rho_1, \rho_2\}$[3] completes this step of the proof.

*Step 5.* We now make a few observations about the behavior of $\Psi^{\iota,\delta_j}(\eta, \xi) := \Psi^{\iota,\delta_j}(\eta, \xi; \gamma')$ and its implications. From now on, we fix $j$ large enough so that all relevant bounds hold. We also drop the indices altogether and assume $\delta := \delta(\iota) \in o(\iota^{-1})$. Recall the function $\Psi^\iota(\eta, \xi) = u(\eta) - v^{\epsilon,\delta}(\xi + \overline{\alpha}) - \frac{1}{2}\iota|\eta - \xi|^2$ achieves a maximum $m_\iota$ at $(\eta_\iota, \xi_\iota) \in \mathfrak{D} \times \mathfrak{D}$. By standard arguments

$$\lim_{\iota \to \infty} \iota |\eta_\iota - \xi_\iota|^2 \to 0$$

and

$$m_\iota \searrow m := \sup_{\eta \in \mathfrak{D}} \{u(\eta) - v^{\epsilon,0}(\eta + \overline{\alpha})\} > 0.$$

*Step 6.* Write

$$F(x, y, w, p, X) := \beta w - rxp^1 - \sum_{i=1}^{d} \mu^i y^i p^{i+1} - \frac{1}{2}\mathrm{Tr}[\sigma(y)\sigma(y)^\top X] - \tilde{U}(p^1),$$

where we write $\sigma(y)\sigma(y)^\top$ instead of $\sigma\sigma^\top \mathbf{D}_{yy}$ to emphasize the $y$ argument more explicitly in the calculations. Since $u$ is a sub-solution and $v^{\epsilon,\delta}$ is a super-solution, the Crandall-Ishii lemma yields

$$(\eta_\iota, u(\eta_\iota), p_\iota, X_\iota) \in \overline{J^{2,+}}u(\eta_\iota)$$

---
[3] Note that while $\gamma'$ increases up to $\hat{\gamma}$, so might $L(\gamma')$ increase to infinity. However, in each case $L(\gamma')$ is finite and all of the inequalities continue to hold for any $\gamma'$ chosen large enough.

and
$$(\xi_\iota + \overline{\alpha}, v^{\epsilon,\delta}(\xi_\iota + \overline{\alpha}), p_\iota, Y_\iota) \in \overline{J^{2,-}}v^{\epsilon,\delta}(\xi_\iota + \overline{\alpha}).$$

Naturally, there corresponds a test function $\varphi_\iota \geq u$ to $(\eta_\iota, u(\eta_\iota), p_\iota, X_\iota) \in \overline{J^{2,+}}u(\eta_\iota)$ which touches $u$ at $\eta_\iota$ and is $C^2$ in a neighborhood of $\eta_\iota$. We claim that $(\varphi_\iota - \mathbf{M}\varphi_\iota)_*(\eta_\iota) \leq 0$ as $\iota \to \infty$. If this is not the case, then we use the fact that $u$ is a sub-solution, $v^{\epsilon,\delta}$ is a super-solution, and that $\xi_\iota + \overline{\alpha} \in \mathring{\mathfrak{D}}$ to obtain

$$0 \geq F(\eta_\iota, u(\eta_\iota), p_\iota, X_\iota) - F(\xi_\iota + \overline{\alpha}, v^{\epsilon,\delta}(\xi_\iota + \overline{\alpha}), p_\iota, Y_\iota)$$
$$= \beta u(\eta_\iota) - rx(\eta_\iota)p_\iota^1 - \sum_{i=1}^d \mu^i y^i(\eta_\iota)p_\iota^{i+1} - \frac{1}{2}\mathrm{Tr}[\sigma(y(\eta_\iota))\sigma(y(\eta_\iota))^\top X_\iota] - \tilde{U}(p_\iota^1)$$
$$\quad - \beta v^{\epsilon,\delta}(\xi_\iota + \overline{\alpha}) + rx(\xi_\iota + \overline{\alpha})p_\iota^1 + \sum_{i=1}^d \mu^i y^i(\xi_\iota + \overline{\alpha})p_\iota^{i+1}$$
$$\quad + \frac{1}{2}\mathrm{Tr}[\sigma(y(\xi_\iota + \overline{\alpha}))\sigma(y(\xi_\iota + \overline{\alpha}))^\top Y_\iota] + \tilde{U}(p_\iota^1)$$
$$\geq \beta(u(\eta_\iota) - v^{\epsilon,\delta}(\xi_\iota + \overline{\alpha})) - \frac{1}{2}\mathrm{Tr}[\sigma(y(\eta_\iota))\sigma(y(\eta_\iota))^\top X_\iota - \sigma(y(\xi_\iota))\sigma(y(\xi_\iota))^\top Y_\iota]$$
$$\geq \beta(u(\eta_\iota) - v^{\epsilon,\delta}(\xi_\iota + \overline{\alpha})) - 3\iota|\eta_\iota - \xi_\iota|^2 L_\sigma$$
$$\geq \frac{1}{2}\beta(u(\eta_\iota) - v^{\epsilon,\delta}(\xi_\iota + \overline{\alpha})) > 0, \quad \text{for all } \iota \text{ sufficiently large,}$$

which yields a contradiction. Thus, $(\varphi_\iota - \mathbf{M}\varphi_\iota)_*(\eta_\iota) \leq 0$ as $\iota \to \infty$.

*Step 7.* We again consider the test function $\varphi_\iota$ from the previous step. In this step, we aim to modify the test function $\varphi_\iota$ so as to obtain a favorable inequality involving $u$. To begin with, note that there exists a net $\{\eta_{\iota,k(\iota)}\}_\iota \subset \mathfrak{D}$ such that $|\eta_\iota - \eta_{\iota,k(\iota)}| = o(\iota^{-1})$ and

$$\varphi_\iota(\eta_\iota) - M\varphi_\iota(\eta_{\iota,k(\iota)}) \leq o(\iota^{-1}).$$

Consequently, $\mathcal{I}(\eta_{\iota,k(\iota)}) \neq \emptyset$. we may modify $\varphi_\iota$ in the following manner. Let $\tilde{r} > 0$ be sufficiently small so that $\varphi_\iota$ is $C^2$ in the ball $B_{\tilde{r}}(\eta_\iota)$ where $\tilde{r} < \frac{1}{2}\max\{\lambda, |\eta_\iota|\}$. Let $R_{\tilde{r}}(\eta_\iota) := \{\eta \in B_{\tilde{r}}(\eta_\iota) : \mathcal{I}(\eta) \neq \emptyset\}$.

Consider the set
$$\mathcal{I}(R_{\tilde{r}}(\eta_\iota)) := \{\zeta + \nu : \zeta \in R_{\tilde{r}}(\eta_\iota), \, \nu \in \mathcal{I}(\zeta)\}.$$

Note that by construction the Hausdorff distance between $R_{\tilde{r}}(\eta_\iota)$ and $\mathcal{I}(R_{\tilde{r}}(\eta_\iota))$ is positive. Let $\{\zeta_t\}_{t\in\mathbb{R}} \subset R_{\tilde{r}}(\eta_\iota)$ be any continuum of points such that $z(\zeta_s) < z(\zeta_t)$ for $s < t$ and $\mathcal{I}(\{\zeta_t\}_{t\in\mathbb{R}}) = \mathcal{I}(R_{\tilde{r}}(\eta_\iota))$. Define the function

$$\hat{\varphi}_\iota(\zeta) = \begin{cases} \varphi_\iota & \text{on } \mathfrak{D} \setminus \mathcal{I}(R_{\tilde{r}}(\eta_\iota)) \\ (\mathbf{M}u)^*(\zeta_t) & \text{on } \mathcal{I}(\zeta_t), \forall t \in \mathbb{R}. \end{cases}$$

Redefining $\varphi_\iota$ so that $\varphi_\iota = \hat{\varphi}_\iota^*$ preserves its local properties at $\eta_\iota$, maintains upper-semicontinuity globally, and ensures $\mathbf{M}\varphi_\iota(\eta_{\iota,k(\iota)}) = (\mathbf{M}\varphi_\iota)^*(\eta_{\iota,k(\iota)}) =$ $(\mathbf{M}u)^*(\eta_{\iota,k(\iota)})$. The first equality follows by the fact that $\mathbf{M}\varphi_\iota$ is upper-semi-continuous (since $\varphi_\iota$ is upper semi-continuous) and the second from the definition of $\varphi_\iota$. This is where the relaxation of test functions in Definition 3.4 is needed.

Therefore $0 \geq (\varphi_\iota(\eta_\iota) - \mathbf{M}\varphi_\iota(\eta_\iota))_* = u(\eta_\iota) - (\mathbf{M}u)^*(\eta_{\iota,k(\iota)})$. We already know $v^{\epsilon,\delta}(\xi_\iota + \overline{\alpha}) - \mathbf{M}v^{\epsilon,\delta}(\xi_\iota + \overline{\alpha}) \geq \frac{1}{2}\epsilon\rho$. Combining these facts with our previous observations yields

$$u(\eta_\iota) - v^{\epsilon,\delta}(\xi_\iota + \overline{\alpha}) < (\mathbf{M}u)^*(\eta_{\iota,k(\iota)}) - \mathbf{M}v(\xi_\iota + \overline{\alpha}) - \frac{1}{2}\epsilon\rho.$$

We may even pass to a maximizing sequence $\{\eta_{\iota,k(\iota),l(\iota)}\}_\iota$ such that $\mathcal{I}(\eta_{\iota,k(\iota),l(\iota)}) \neq \emptyset$ and $\mathbf{M}u(\eta_{\iota,k(\iota),l(\iota)}) = (\mathbf{M}u)^*(\eta_{\iota,k(\iota)}) + o(\iota^{-1})$ and $\eta_{\iota,k(\iota),l(\iota)} = \eta_\iota + o(\iota^{-1})$.

Note that $|\eta_\iota - \xi_\iota| \to 0$ implies that for $\iota$ sufficiently large $\eta_{\iota,k(\iota),l(\iota)} \leq \xi_\iota + \overline{\alpha}$. Therefore, $\mathcal{I}(\xi_\iota + \overline{\alpha}) \neq \emptyset$. It follows that $\mathbf{M}v(\xi_\iota + \overline{\alpha}) > -\infty$.

*Step 8.* All the ingredients are now present to derive the desired contradiction. As we have already seen

$$u(\eta_\iota) \leq \mathbf{M}u(\eta_{\iota,k(\iota),l(\iota)}) + o(\iota^{-1})$$

and

$$v^{\epsilon,\delta}(\xi_\iota + \overline{\alpha}) - \mathbf{M}v^{\epsilon,\delta}(\xi_\iota + \overline{\alpha}) \geq \frac{\epsilon\rho}{2}.$$

We proceed to write

$$\begin{aligned}
m_\iota &= u(\eta_\iota) - v^{\epsilon,\delta}(\xi_\iota + \overline{\alpha}) - \frac{1}{2}\iota|\eta_\iota - \xi_\iota|^2 \\
&\leq \mathbf{M}u(\eta_{\iota,k(\iota),l(\iota)}) - \mathbf{M}v^{\epsilon,\delta}(\xi_\iota + \overline{\alpha}) - \frac{\epsilon\rho}{2} - \frac{1}{2}\iota|\eta_\iota - \xi_\iota|^2 + o(\iota^{-1}) \\
&\leq \sup_{\nu \in \mathcal{I}(\eta_{\iota,k(\iota),l(\iota)})} \{u(\eta_{\iota,k(\iota),l(\iota)} + \nu) - v^{\epsilon,\delta}(\xi_\iota + \overline{\alpha} + \nu)\} - \frac{1}{2}\iota|\eta_\iota - \xi_\iota|^2 - \frac{\epsilon\rho}{2} + o(\iota^{-1}) \\
&= \sup_{\nu \in \mathcal{I}(\eta_\iota^{k_\iota})} \{u(\eta_{\iota,k(\iota),l(\iota)} + \nu) - v^{\epsilon,\delta}(\xi_\iota + \overline{\alpha}\nu) - \frac{1}{2}\iota|\eta_{\iota,k(\iota),l(\iota)} - \xi_\iota|^2\} - \frac{\epsilon\rho}{2} \\
&\quad + \frac{1}{2}\iota|\eta_{\iota,k(\iota),l(\iota)} - \xi_\iota|^2 - \frac{1}{2}\iota|\eta_\iota - \xi_\iota|^2 + o(\iota^{-1}) \\
&\leq m_\iota + \frac{1}{2}\iota(|\eta_{\iota,k(\iota),l(\iota)} - \xi_\iota|^2 - |\eta_\iota - \xi_\iota|^2) + o(\iota^{-1}) - \frac{\epsilon\rho}{2} \\
&\leq m_\iota - \frac{\epsilon\rho}{2} + o(1),
\end{aligned}$$

which yields the contradiction. $\square$

## 5 Numerical results

In this section, we construct and implement algorithms to determine the no-trade region and optimal strategies at the leading order. The approximate solutions are obtained by homogenization as $\lambda_f, \lambda_p \to 0$. This technique has been applied successfully to the study of the Merton problem under various frictions. The method entails expanding the frictional value function as a series in $\lambda_f, \lambda_p$ around the frictionless Merton value function. The technique requires that the frictionless value function is finite (this is a standing assumption).

One of the virtues of analysing the asymptotic problem numerically, apart from reduction of dimensionality, is the fact that the no-trade region is bounded in each iso-wealth plane. Therefore, if this region is contained in the computation domain, the control will be active at the boundary, rendering the boundary conditions immaterial.

## 5.1 The frictionless problem

We begin by considering the problem of investing and consuming in a setting where there are no transaction costs. Since trades are then costless, the corresponding value function does not depend separately on the positions $x, y$ in the safe and the risky assets, but only on total wealth $z = x + y \cdot \mathbf{1}_d$. As is well known (cf., e.g., [17, Chapter X]), the frictionless value function solves the dynamic programming equation

$$0 = \widetilde{U}\bigl(v_z(z)\bigr) - \beta v(z) + \mathscr{L}_0 v(z), \tag{5.1}$$

where

$$\mathscr{L}_0 v(z) = v_z(z) z r + v_z(z)(\mu - r\mathbf{1}_d) \cdot \theta(z) + \frac{1}{2} v_{zz}(z) |\sigma^\top \theta(z)|^2, \tag{5.2}$$

and the corresponding optimal consumption rate and optimal risky positions are given by

$$\kappa(z) = (U')^{-1}\bigl(v_z(z)\bigr) \tag{5.3}$$

and

$$\theta(z) := -\frac{v_z(z)}{v_{zz}(z)} (\sigma\sigma^\top)^{-1} (\mu - r\mathbf{1}_d). \tag{5.4}$$

## 5.2 Homogenization

Before embarking on a full asymptotic analysis, it will be convenient at this stage to rewrite $\lambda_f = \epsilon^4$ and $\lambda_p = \nu_p \epsilon^3$. We choose this parametrization so that all the forthcoming expansions contain only integral powers of $\epsilon$. We will also abuse notation and write $v^\epsilon := v^\lambda$.

Define the *fast-variable* in the expansion as

$$\xi = \frac{y - \theta(z)}{\epsilon},$$

Since we perform an inner expansion in the no-trade region, we scaled $\xi$ by the width of the no-trade region. Standard heuristic arguments can be applied to guess that width should be on the order of $\epsilon = \lambda_f^{1/4} \propto \lambda_p^{1/3}$. An informed guess yields the ansatz

$$v^\epsilon(x, y) = v(z) - \epsilon^2 u(z) - \epsilon^4 w(z, \xi) + o(\epsilon^3).$$

Formally substituting the ansatz into the DPE gives rise to the so-called *corrector equations* for $u$ and $w$. The DPE is comprised of two parts, an elliptic part and a non-local part and they require separate expansions.

### 5.2.1 Expansion in no-trade region

The elliptic expression has already been approximated to leading order in a number of papers, e.g. [1]. The same computation yields

$$\beta v^\epsilon(x,y) - \widetilde{U}\big(v^\epsilon_x(x,y)\big) - \mathscr{L}v^\epsilon(x,y)$$
$$= -\epsilon^2\Big(\beta u(z) - \mathscr{L}_0 u(z) + \kappa(z)u_z(z) + \frac{|\sigma^\top \xi|^2}{2}v_{zz}(z)$$
$$- \frac{1}{2}\mathrm{Tr}[\alpha(z)\alpha(z)^\top w_{\xi\xi}(z,\xi)]\Big) + o(\epsilon^2),$$

for the differential operator $\mathscr{L}_0$ from (5.2) and

$$\alpha(z) = (I_d - \theta_z(z)\mathbf{1}_d^\top)\mathrm{diag}\,[\theta(z)]\,\sigma.$$

Satisfying the elliptic part of equation (2.4) between bulk trades—at the leading order $O(\epsilon^2)$—is therefore tantamount to

$$0 = \beta u(z) - \mathscr{L}_0 u(z) + \kappa(z)u_z(z) + \frac{|\sigma^\top \xi|^2}{2}v_{zz}(z) - \frac{1}{2}\mathrm{Tr}[\alpha(z)\alpha(z)^\top w_{\xi\xi}(z,\xi)]. \quad (5.5)$$

### 5.2.2 Expansion in trade region

By definition, $v^\epsilon \geq \mathbf{M}v^\epsilon$ holds at all points of the domain. Inserting the ansatz, this reads

$$v(z) - \epsilon^2 u(z) - \epsilon^4 w(z,\xi) + o(\epsilon^3) \geq \sup_{\hat{\xi}}\left\{v(\hat{z}) - \epsilon^2 u(\hat{z}) - \epsilon^4 w(\hat{z},\hat{\xi})\right\},$$

where

$$\hat{z} = z - \epsilon^4 - \nu_p\epsilon^3\|\hat{y} - y\|_1 = z - \epsilon^4(1 + \nu_p\|\hat{\xi} - \xi\|_1)$$

and where the supremum is taken over admissible portfolio positions $\hat{\xi}$ with wealth $\hat{z}$. If $w$ is smooth, then $w(\hat{z},\hat{\xi}) = w(z,\hat{\xi}) + o(\epsilon^3)$. Proceeding formally, we observe

$$0 \geq \sup_{\hat{\xi}}\left\{v(\hat{z}) - v(z) - \epsilon^2(u(\hat{z}) - u(z)) - \epsilon^4(w(\hat{z},\hat{\xi}) - w(z,\xi))\right\}$$
$$= \epsilon^4\sup_{\hat{\xi}}\{-v_z(z)(1 + \nu_p\|\hat{\xi} - \xi\|_1) + w(z,\xi) - w(z,\hat{\xi})\} + o(\epsilon^4)$$
$$= \epsilon^4\left(w(z,\xi) - v_z(z) - \inf_{\hat{\xi}}\{v_z(z)\nu_p\|\hat{\xi} - \xi\|_1 + w(z,\hat{\xi})\}\right) + o(\epsilon^4).$$

Therefore, at the leading order $w$ should satisfy

$$w(z,\xi) \leq v_z(z) + \inf_{\xi'}\{v_z(z)\nu_p\|\xi' - \xi\|_1 + w(z,\xi')\} \tag{5.6}$$

with equality holding in the trade region.

**Remark 5.1.** *The condition* (5.6) *has an appealing interpretation. It was shown in previous studies that the function $w$ can be viewed as the potential in an ergodic control problem. Since the cost of trading from $\xi$ to $\hat{\xi}$ is $(1 + \nu_p\|\hat{\xi} - \xi\|_1)\epsilon^4$, the leading order loss of utility from trading is*

$$v(z) - v(\hat{z}) \approx v_z(z)(z - z') = v_z(z)(1 + \nu_p\|\hat{\xi} - \xi\|_1)\epsilon^4.$$

*Thus,* (5.6) *implies that trades should occur precisely when the utility loss can be offset by the change in potential $w(z,\xi') - w(z,\xi)$.* □

### 5.2.3 Corrector equations

Rewriting the DPE (2.4) using the formal expansion results yields a pair of coupled equations called corrector equations. Given any $z > 0$, we wish to find an unknown pair $(a(z), w(z,\cdot)) \in R^+ \times C^2(\mathbb{R}^d)$ which satisfies the *first corrector equation*

$$\max\left\{\frac{1}{2}|\sigma^\top\xi|^2 v_{zz}(z) - \frac{1}{2}\text{Tr}\left[\alpha(z)\alpha(z)^\top w_{\xi\xi}(z,\xi)\right] + a(z),\right. \tag{5.7}$$

$$\left. w(z,\xi) - v_z(z) - \inf_{\hat{\xi}\in\mathbb{R}^d}\left\{v_z(z)\nu_p\|\hat{\xi} - \xi\|_1 + w(z,\hat{\xi})\right\}\right\} = 0. \tag{5.8}$$

A function $a(z)$ is then obtained from the first corrector equation by solving for every $z > 0$. The leading order term in the expansion of $v^\epsilon$ is obtained by solving for $u(z)$ in the *second corrector equation*

$$\beta u(z) - \mathcal{L}_0 u(z) + \kappa(z)u(z) = a(z), \quad z \in \mathbb{R}_+.$$

## 5.3 Numerical method

As was alluded to earlier, the first corrector equation (5.7) arises from an ergodic control problem. To be more precise, the problem is defined as follows. Find $a(z)$ given by

$$a(z) := \inf_{m,\tau} J(z,m,\tau)$$

where the cost functional $J$ is defined by

$$J(z,m,\tau) := \limsup_{T\to\infty} \frac{1}{T}\mathbb{E}\left[\int_0^T -v_{zz}(z)\frac{|\sigma^\top\xi_s|^2}{2}ds \right.$$
$$\left. + v_z(z)\sum_{k=1}^\infty (1 + \nu_p\|m\|_1)1_{\{\tau_k \leq T\}}\right],$$

and the state process $\xi = (\xi_t)_{t \in [0,\infty)}$ evolves according to

$$\xi_t = \xi_0 + \alpha(z) B_t + \sum_{k=1}^{\infty} m_k 1_{\{\tau_k \leq t\}}$$

for a standard $d$-dimensional Brownian motion $B$.

In a setting where the control is not of impulse type, it can be approximated with controls which are absolutely continuous with respect to time. The problem can then be discretized in a way which makes standard policy iteration techniques apply. In fact, if done appropriately, the discrete HJB equation will be a penalized version of the singular control counterpart. This is done implicitly in [38]. Similarly, the goal in our setting is to approximate the impulse controls in such a way that policy iteration gives the solution to a penalized version of the first corrector equation.

The problem will be solved for fixed $z$, and as a starting point, the problem is discretized in the remaining space variables as described in [8]. Any discretization not explicitly stated here will follow that scheme. The impulse control is discretized separately as described below, and the resulting controlled process is then a continuous-time Markov chain for which a policy iteration scheme can be implemented.

We denote by $\mathcal{L}^m$ the discretized infinitesimal generator of $\xi$ on some grid, corresponding to the feedback control $m = (m^1, \ldots, m^d)$. Here $m^i$ denotes the distance to move in direction $i$, i.e., the number of grid points by which to move times the mesh width in that dimension. Then $\mathcal{L}^m(\xi, \xi')$ is the transition rate from $\xi$ to $\xi'$. Let $\mathcal{L}^m = \mathcal{L} + \mathcal{L}_K^m$, where $\mathcal{L}$ is the operator corresponds to the elliptic part of the equation, and $\mathcal{L}_K^m$ consists of additional terms arising when a control is active. To account for the impulse control jumps, we set

$$\mathcal{L}_K^m(\xi, \xi) = -K \text{ and } \mathcal{L}_K^m(\xi, \xi + m) = K$$

for some suitably large value of $K$. This is the operator of the ergodic control problem above, but where jumps in the direction of the control do not happen with certainty. Instead, conditioned on a jump occurring, the state moves in the direction of the control with some probability which tends to 1 as $K \to \infty$. Moreover, the probability making a jump in a given time interval also tends to 1 as $K \to \infty$ if the control is active. In this sense the $K$ approximates the singularity of the control.

Finally, by also discretizing the running cost, the problem is reformulated as a continuous time Markov decision process for which standard policy iteration techniques apply (cf., e.g., [40]). That is, compute

$$a(z) := \inf_m J(z, m),$$

where $J$ is the cost functional given by

$$J(z,m) := \limsup_{T\to\infty} \frac{1}{T}\mathbb{E}\left[\int_0^T -v_{zz}(z)\frac{|\sigma^\top \xi_s|^2}{2} + Kv_z(z)(1+\nu_p\|m\|_1)1_{\{m\neq 0\}}ds\right],$$

subject to

$$\xi_t = \xi_0 + \hat{B}_t^\alpha + \sum_{s\leq t} m(\xi_s)\Delta N_s$$

for a Poisson process $N$ with rate $K$ and where $\hat{B}_t^\alpha$ is the spatial discretization of the process $\alpha(z)B_t$ corresponding to $\mathcal{L}$. The DPE for this ergodic control problem is precisely a discretization of the elliptic part of the first corrector equation together with a penalization term for the trade condition, where the penalty is $K$.

For a fixed wealth $z$, let $\mathcal{D}$ be a finite mesh on $\mathbb{R}^d$, where $d$ denotes the number of risky assets. Denote by $N$ the number of grid points in $\mathcal{D}$ and set

$$f^m(\xi) := -v_{zz}(z)\frac{|\sigma^\top \xi|^2}{2} + Kv_z(z)(1+\nu_p\|m\|_1)1_{\{m\neq 0\}}, \quad \forall \xi \in \mathcal{D}.$$

Naturally, since the computational domain $\mathcal{D}$ is finite, we must also specify boundary conditions. If the probability of a controlled process reaching the boundary is small enough, the boundary conditions have a negligible effect on the final outcome. In fact, as mentioned above, if the no-trade region is contained in the domain, and therefore the control active at the boundary, the effect of the boundary conditions is even further diminished. To this end, $\mathcal{D}$ should be sufficiently large that the continuation region is entirely contained in the interior of $\mathcal{D}$. That is, at all points on $\partial \mathcal{D}$, the investor should want to trade into the interior of $\mathcal{D}$. The boundary conditions are then chosen so that $\mathcal{L}^m$ can still be interpreted as a transition rate matrix (e.g. a homogeneous Neumann condition).

The algorithm proceeds as follows. Initialize a starting policy $m_0 \in \mathbb{R}^{d\times N}$ and choose a starting value $a_0 > 0$ sufficiently large. Then select a tolerance level,[4] $\tau \geq 0$ and iterate:

---

[4] We can even take $\tau = 0$ since eventually $a_i = a_{i+1}$.

**Policy iteration algorithm**

(i) Compute[5] $(w, a_{i+1}) \in \mathbb{R}^N \times \mathbb{R}^+$ such that

$$\sum_{\xi' \in \mathcal{D}} \mathcal{L}^{m_i}(\xi, \xi') w(\xi') + f^{m_i}(\xi) = a_{i+1}, \quad \forall \xi \in \mathcal{D}.$$

Halt if $|a_i - a_{i+1}| \leq \tau$.

(ii) For each $\xi \in \mathcal{D}$, compute $m_{i+1}(\xi)$ where

$$m_{i+1}(\xi) \in \underset{\xi + \hat{m} \in \mathcal{D}}{\arg\min} \left( \sum_{\xi' \in \mathcal{D}} \mathcal{L}^{\hat{m}}(\xi, \xi') w(\xi') + f^{\hat{m}}(\xi) \right).$$

(iii) Return to step (i).

## 5.4 Interpretation of results

Henceforth, the number of assets will be either 1 or 2.

In the results below we will use the following notation unless otherwise stated. The returns of the assets are given by $\mu_1$ and $\mu_2$, i.e.,

$$\mu = \begin{bmatrix} \mu_1 \\ \mu_2 \end{bmatrix}.$$

The covariance matrix is defined through the variances $\sigma_1$, $\sigma_2$, and the correlation $\rho$ by

$$\sigma\sigma^\top = \begin{bmatrix} \sigma_1^2 & \rho\sigma_1\sigma_2 \\ \rho\sigma_1\sigma_2 & \sigma_2 \end{bmatrix}.$$

As in the previous sections, $r$ is the interest rate, $\beta$ denotes the impatience rate, and $\gamma$ is the elasticity of the CRRA utility function

$$U(c) = \begin{cases} \frac{c^{1-\gamma}}{1-\gamma}, & \gamma \neq 0, \\ \ln c, & \gamma = 0. \end{cases}$$

The asymptotic problem is solved in the $\xi$-space for fixed $z$. To make a meaningful interpretation of the result, we think of this as the solution in $\xi$-space for some non-zero level of $\epsilon > 0$. With this interpretation, the transaction costs are precisely $\lambda_f = \epsilon^4$ and $\lambda_p = \nu_p \epsilon^3$. The value of $\epsilon$ only enters in the interpretation and the choice only affects the relationship between the two transaction costs and the wealth level. To simplify interpretation it is chosen so that $\lambda_f = 1$ and $\lambda_p = \nu_p$. The relative deviation is given by $\xi/z$.

---

[5]The boundary conditions on $\mathcal{D}$ are chosen such that $\sum_{\xi' \in \mathcal{D}} \mathcal{L}^m(\xi, \xi') = 0$ for all $\xi \in \mathcal{D}$. Since the system is *underdetermined*, $w^j$ can be normalized so that $w(0) = 0$ without modifying the equations.

Finally, the two-dimensional figures are cuts in planes where the wealth level $z$ is fixed. The colors carry information about how to optimally trade, and their meanings are given in the table below.

| | | |
|---:|:---:|:---|
| white |  | no trading |
| yellow |  | trading in asset 1 |
| red |  | trading in asset 2 |
| orange |  | trading in both assets |

Whether buying or selling is optimal is not indicated by the color since the optimal strategy always moves the portfolio closer to the Merton proportion. This is visualized by the black points, which are the points to which the investor chooses to trade.

## 5.5 Benchmark testing

Since the DPE of the discrete problem is the equation of a penalized version of the impulse control problem, we expect that solutions will converge to those of the first corrector equation. We will not undertake a rigorous convergence analysis here. Instead, we demonstrate that solutions computed by the scheme agree with available analytical benchmarks. These benchmarks include: fixed and proportional costs in one dimension as well as only fixed costs in two dimensions.

In one dimension with both fixed and proportional transaction costs it is possible to find an analytic characterization of the first corrector equation. This is accomplished by using a smooth fit condition between a fourth order polynomial inside the trade region and linear growth outside, and then solve for the polynomial coefficients as well as the free boundary. A comparison between this analytic solution and the policy iteration result is found in Figure 3. The two solutions coincide to any extent discernible.

In the two-dimensional setting, the trading strategy is analytically known in the absence of proportional transaction costs (cf. [1]). In Figure 4, the analytic free boundary is plotted on top of the iterative solution and the fit is near perfect. Moreover, the analytic solution tells us that all trades are to the Merton proportion, and we verify that this is indeed also the result from policy iteration.

Note that the long-term distribution in the second case does not coincide with any of the main axes of the ellipse describing the no-trade region. We attribute this to the fact that the risk aversion and market parameters induce a frictionless trader to keep 8.6 times more of the wealth in asset one (vertical axis) than in asset two (horizontal axis). Thus, most of the fluctuation of the portfolio position will be in the corresponding direction. This information is incorporated in the matrix $\alpha^\top \alpha$. Indeed,

$$\alpha^\top \alpha \approx \begin{bmatrix} 42607.9 & 669.990 \\ 669.990 & 147.787 \end{bmatrix}.$$

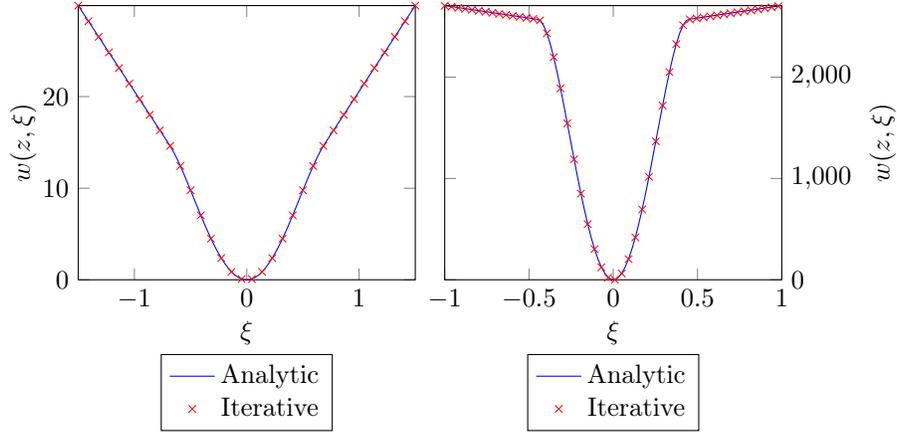

Figure 3: Known analytic solution plotted alongside the policy iteration solution for two different parameter configurations.

Hence, the volatility of $\xi^1$ is around 288 times larger than that of $\xi^2$, and the correlation is approximately $10^{-4}$.

## 5.6 Proportional transaction costs

In the setting of only proportional transaction costs, Figure 5 shows the boundaries of the trade regions with changing covariance matrix. For comparison with [36], we here define the covariance matrix through

$$\sigma = \begin{bmatrix} 0.4 - \kappa & \kappa \\ \kappa & 0.4 - \kappa \end{bmatrix},$$

where $\kappa$ is $-0.1$, $-0.05$, $0$, $0.025$, $0.05$ and $0.075$. Here the interpretation of $\epsilon$ is such that $\lambda_p = 0.02$, i.e., such that we have proportional transaction costs of 2 %. Note that due to homotheticity exhibited by the proportional cost problem, this only affects the size of the region, since the relative deviation is independent of the wealth level. Observe that the regions are not convex.

## 5.7 Mixed transaction costs

Introducing an additional transaction cost to either of the above problems induces a situation where the investor has to balance the proportional cost associated with trading distance with the fixed cost induced by initiating a trade. This leads to an optimal strategy characterized by a trading boundary and a target curve. When the portfolio exits the no-trade region, i.e., reaches the trading boundary, the optimal action is to rebalance the portfolio such that the new position is at the target curve.

Examples of such strategies are represented in Figure 6. In particular, the shape resembles the shape with only proportional costs, but with the corners rounded, reminiscent of the fixed cost problem. The selection of figures display how the no-trade region changes with variations in the market parameters, all other things being equal. The qualitative effect of such variations is very similar to in the proportional transaction cost problem.

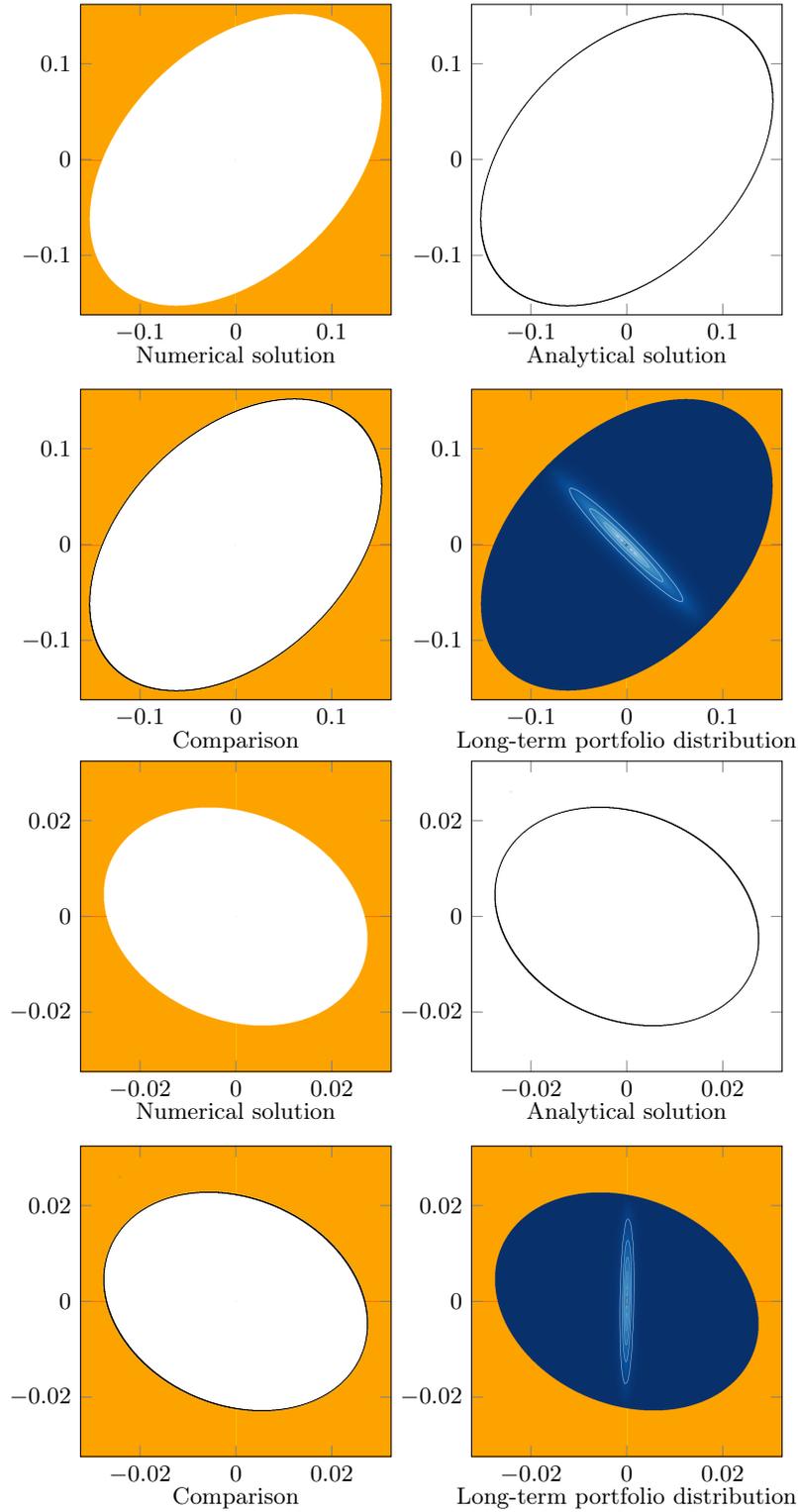

Figure 4: Two comparisons of numerical and analytical solutions, plotted as deviations from the Merton proportions in percentages of wealth held in the risky assets. The blue centers are heatmaps (and level sets) of the (simulated) long-term distribution of portfolio positions.

| rel. pos. | $\beta$ | $\gamma$ | $r$ | $\mu_1$ | $\mu_2$ | $\sigma_1$ | $\sigma_2$ | $\rho$ | $\lambda_f$ | $\lambda_p$ | $z$ |
|---|---|---|---|---|---|---|---|---|---|---|---|
| Top | 1 | 2 | 0.03 | 0.08 | 0.08 | 0.4 | 0.4 | -0.75 | $ 1 | 0% | $ 12'345.67 |
| Bottom | 1 | 7 | 0.03 | 0.08 | 0.04 | 0.4 | 0.2 | 0.35 | $ 1 | 0% | $ 12'345.67 |

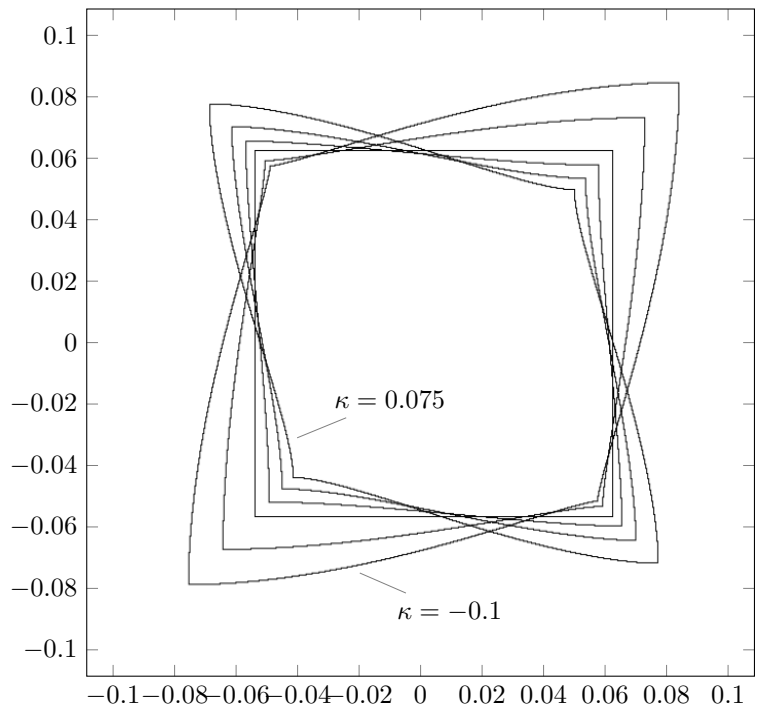

Figure 5: Trade region boundaries with purely proportional transaction costs as the correlation parameter $\kappa$ attains the values $-0.1$, $-0.05$, $0$, $0.025$, $0.05$ and $0.075$. The risk free rate is 3%, $\mu = (0.08, 0.08)$, $\gamma = 2$, and the transaction cost is 2 %.

At low wealth levels, fixed transaction costs are the predominant cost consideration whereas at higher wealth levels proportional costs prevail. Indeed, the results of our computational scheme exhibit these same phenomena. Figures 7 and 8 show the no-trade region as wealth varies *ceteris paribus*. In the 2-d case, the no-trade region appears to interpolate between the elliptic shape associated to fixed costs at low wealth levels and the parallelogram shape associated to proportional costs at high wealth levels. Note also that the total transaction costs are comparatively smaller for a wealthier agent, resulting in smaller no-trade regions in relation to wealth. The 1-d plots in Figure 8 illustrate precisely the $z$ dependence of the optimal trigger and optimal restart barriers.

The results indicate that when proportional costs are present, the no-trade region is not always convex. Indeed, Figure 7 suggests that non-convexity can persist at all wealth levels. This is in stark contrast to the case where there are only fixed costs and the corresponding approximate no-trade region is an ellipse at every level of wealth.

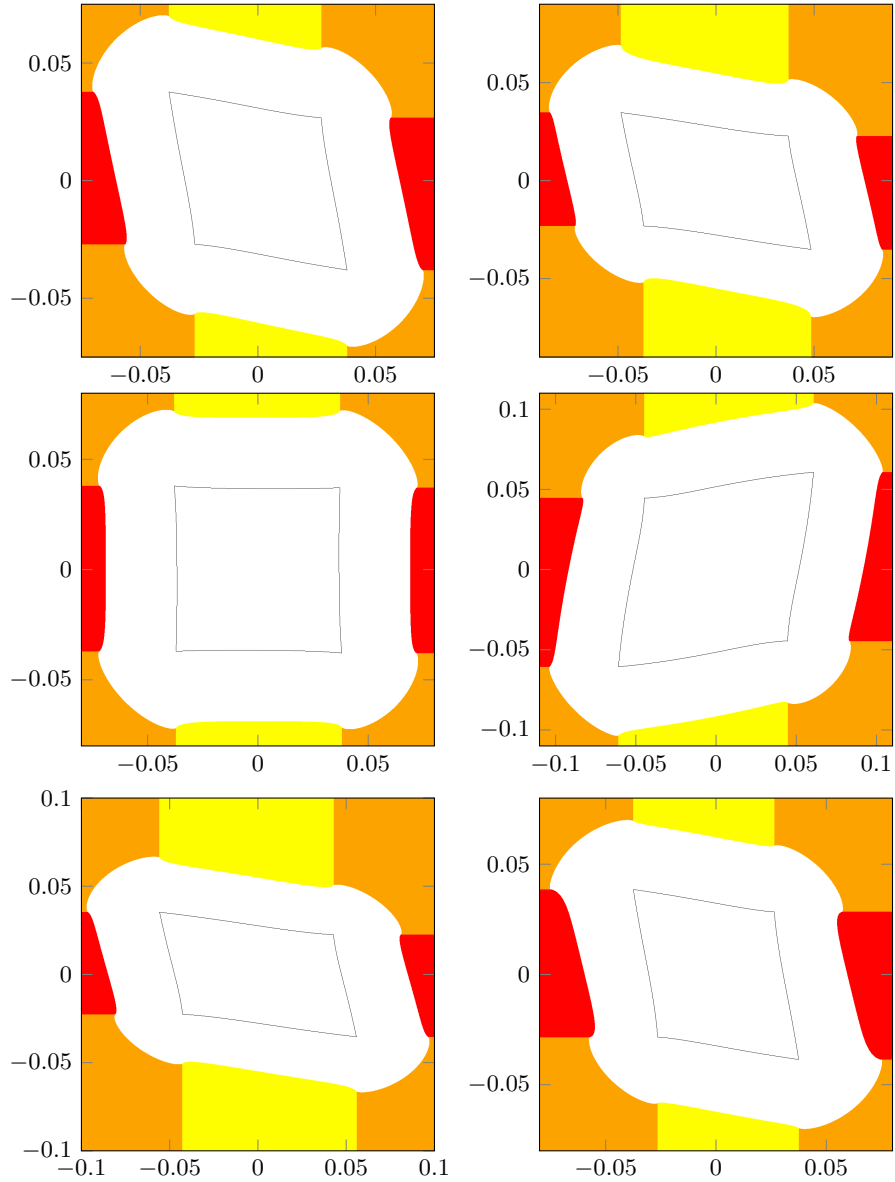

Figure 6: Market parameter dependence as deviations in percentages of wealth.

| rel. pos. | $\beta$ | $\gamma$ | $r$ | $\mu_1$ | $\mu_2$ | $\sigma_1$ | $\sigma_2$ | $\rho$ | $\lambda_f$ | $\lambda_p$ | $z$ |
|---|---|---|---|---|---|---|---|---|---|---|---|
| Top left | 1 | 3 | 0.03 | 0.08 | 0.08 | 0.4 | 0.4 | 0.30 | $ 1 | 3% | $ 10'000 |
| Top right | 1 | 3 | 0.03 | 0.08 | 0.10 | 0.4 | 0.4 | 0.30 | $ 1 | 3% | $ 10'000 |
| Middle left | 1 | 3 | 0.03 | 0.08 | 0.08 | 0.4 | 0.4 | 0 | $ 1 | 3% | $ 10'000 |
| Middle right | 1 | 3 | 0.03 | 0.08 | 0.08 | 0.4 | 0.4 | -0.30 | $ 1 | 3% | $ 10'000 |
| Bottom left | 1 | 3 | 0.03 | 0.08 | 0.08 | 0.4 | 0.3 | 0.30 | $ 1 | 3% | $ 10'000 |
| Bottom right | 1 | 3 | 0.03 | 0.08 | 0.06 | 0.4 | 0.3 | 0.30 | $ 1 | 3% | $ 10'000 |

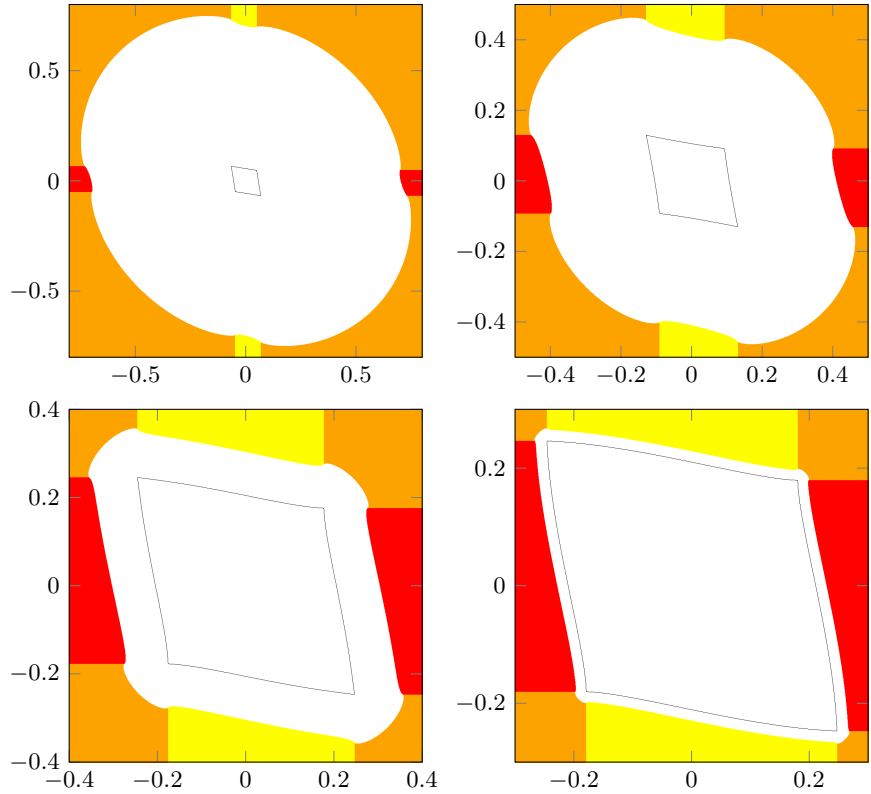

Figure 7: Strategy wealth dependence described as deviations from the Merton portfolio, in units of wealth. The series of figures shows the transition between the two regimes of cost structures. A comparison to Figure 4 and Figure 5 emphasizes the resemblance to purely fixed and purely proportional costs in the extreme cases of wealth.

| rel. pos. | $\beta$ | $\gamma$ | $r$ | $\mu_1$ | $\mu_2$ | $\sigma_1$ | $\sigma_2$ | $\rho$ | $\lambda_f$ | $\lambda_p$ | $z$ |
|---|---|---|---|---|---|---|---|---|---|---|---|
| Top left | 1 | 3 | 0.03 | 0.08 | 0.08 | 0.4 | 0.4 | 0.30 | $ 1 | 3% | $ 100 |
| Top right | 1 | 3 | 0.03 | 0.08 | 0.08 | 0.4 | 0.4 | 0.30 | $ 1 | 3% | $ 1'000 |
| Bottom left | 1 | 3 | 0.03 | 0.08 | 0.08 | 0.4 | 0.4 | 0.30 | $ 1 | 3% | $ 50'000 |
| Bottom right | 1 | 3 | 0.03 | 0.08 | 0.08 | 0.4 | 0.4 | 0.30 | $ 1 | 3% | $ 5'000'000 |

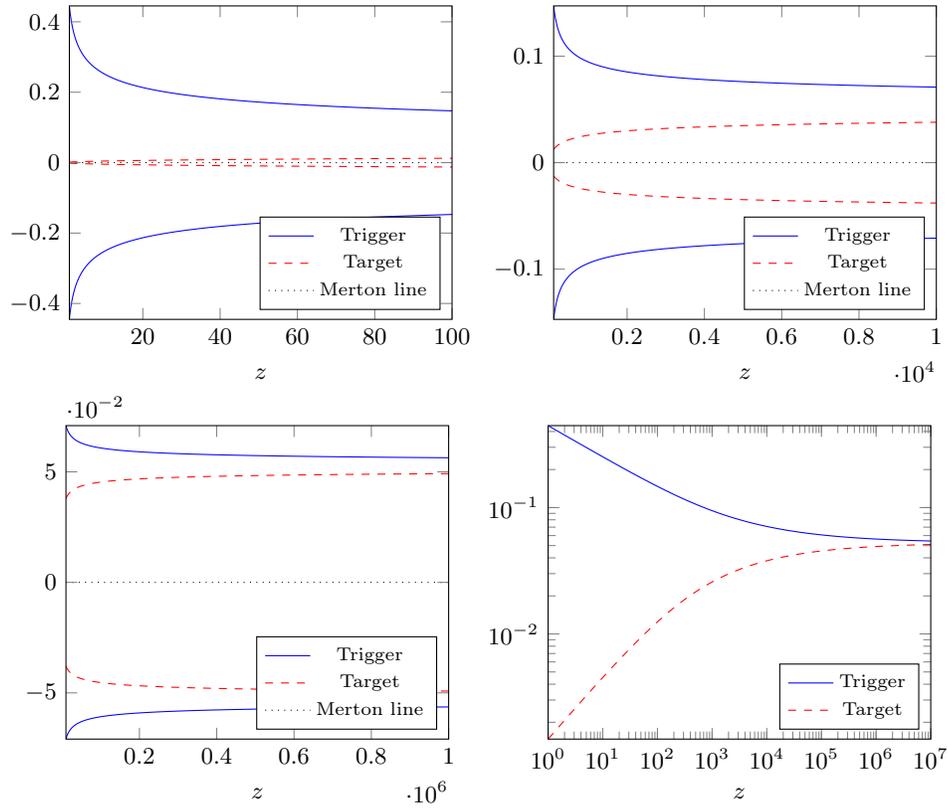

Figure 8: The plots illustrate the wealth dependence of the optimal trading boundary and trading targets as deviations from the Merton portfolio in units of wealth. The converging lines in the bottom right figure clearly show the transition from a fixed cost regime for low wealth to the proportional cost regime as wealth increases.

| $\beta$ | $\gamma$ | $r$ | $\sigma$ | $\mu$ | $\lambda_f$ | $\lambda_p$ |
|---|---|---|---|---|---|---|
| 1 | 5 | 0.01 | 0.2 | 0.04 | $ 1 | 3 % |

## 5.8  Two fixed costs with two risky assets

Appropriately modifying the cost structure in the computational scheme enables us to experiment with more general scenarios. For example, we now wish to examine the no-trade region and optimal strategies when multiple fixed costs are levied for trading *sets* of assets. In the following examples, a fixed transaction cost of $1 is paid for each one of the risky positions rebalanced as opposed to paying a flat fee for re-balancing an entire portfolio of assets. As a result, we are able to not only consider a more realistic scenario in which fixed costs are paid per asset, but we also gain a richer understanding of the interaction between the different cost structures and how that is manifested in the optimal strategies.

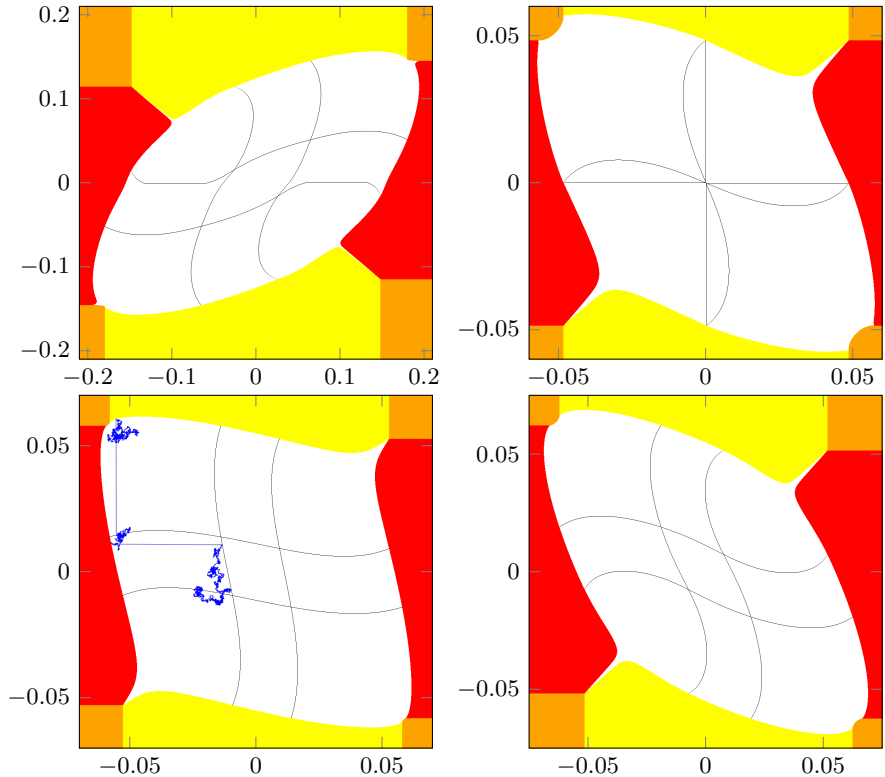

Figure 9: Trading two assets incurs two fixed costs, splitting up the target boundary depending on direction of trades. The blue path is a simulation of a portfolio starting in the upper left corner, visualizing the optimal strategy.

| rel. pos. | $\beta$ | $\gamma$ | $r$ | $\mu_1$ | $\mu_2$ | $\sigma_1$ | $\sigma_2$ | $\rho$ | $\lambda_f$ | $\lambda_p$ | $z$ |
|---|---|---|---|---|---|---|---|---|---|---|---|
| Top left | 1 | 5 | 0.03 | 0.08 | 0.07 | 0.4 | 0.3 | -0.65 | $ 1 | 1.5% | $ 1'000 |
| Top right | 1 | 5 | 0.03 | 0.08 | 0.08 | 0.4 | 0.4 | 0.50 | $ 1 | 0% | $ 1'000 |
| Bottom left | 1 | 5 | 0.03 | 0.08 | 0.08 | 0.4 | 0.4 | 0.30 | $ 1 | 3% | $ 1'000 |
| Bottom right | 1 | 5 | 0.03 | 0.08 | 0.08 | 0.4 | 0.4 | 0.60 | $ 1 | 3% | $ 1'000 |

A clear departure from the previous results is the fact that the inner(restart) region is composed of two pairs of lines in the case $\lambda_p > 0$ which eventually entirely overlap when $\lambda_p = 0$. Upon close inspection, the mapping of the outer boundary to the inner boundary becomes perfectly clear. The yellow region maps orthogonally to the closest horizontal line and the red region maps orthogonally to the closest vertical line. Finally, orange maps to the closest point of intersection of the pairs of lines. The relative widths behave intuitively, as before, according to the correlation and volatilities of the assets. One of the most interesting phenomena is the pinching of the no-trade region that occurs around the $\pm(\xi_1 + \xi_2)$ directions in the case of positive correlation and $\pm(\xi_1 - \xi_2)$ directions in the case of negative correlation (although not illustrated here, there is no such phenomena in the uncorrelated case and the region is perfectly symmetric across the horizontal and vertical axes). Let us take for example the extreme looking case in the bottom right image of Figure 9. In this case, the investor would rather allow their portfolio to deviate a lot further along the $\xi_1 + \xi_2$ axis through the optimal Merton proportion (than perhaps one might expect) than to be transacting in just one (or both) of the assets. Moreover, a slight deviation off that axis seems to incur an additional quadratic cost loss as it immediately leads the investor to trade in one of the two assets.

## 5.9 Long-term distribution

By simulating the excursions of the stocks in $\xi$-space and following the optimal strategy found above, it is possible to generate the long term distributions of portfolio positions. In Figure 10 these distributions are pictured as heatmaps where white means more time is spent at a point, and blue means less. The target positions, i.e., the positions to which we choose to trade, previously drawn as black lines, shine brighter and are clearly distinguishable. As expected, more time is spent alone the shorter axis due to the positive correlation. This is especially evident in Figure 4 where the correlation parameter $\rho$ is very large.

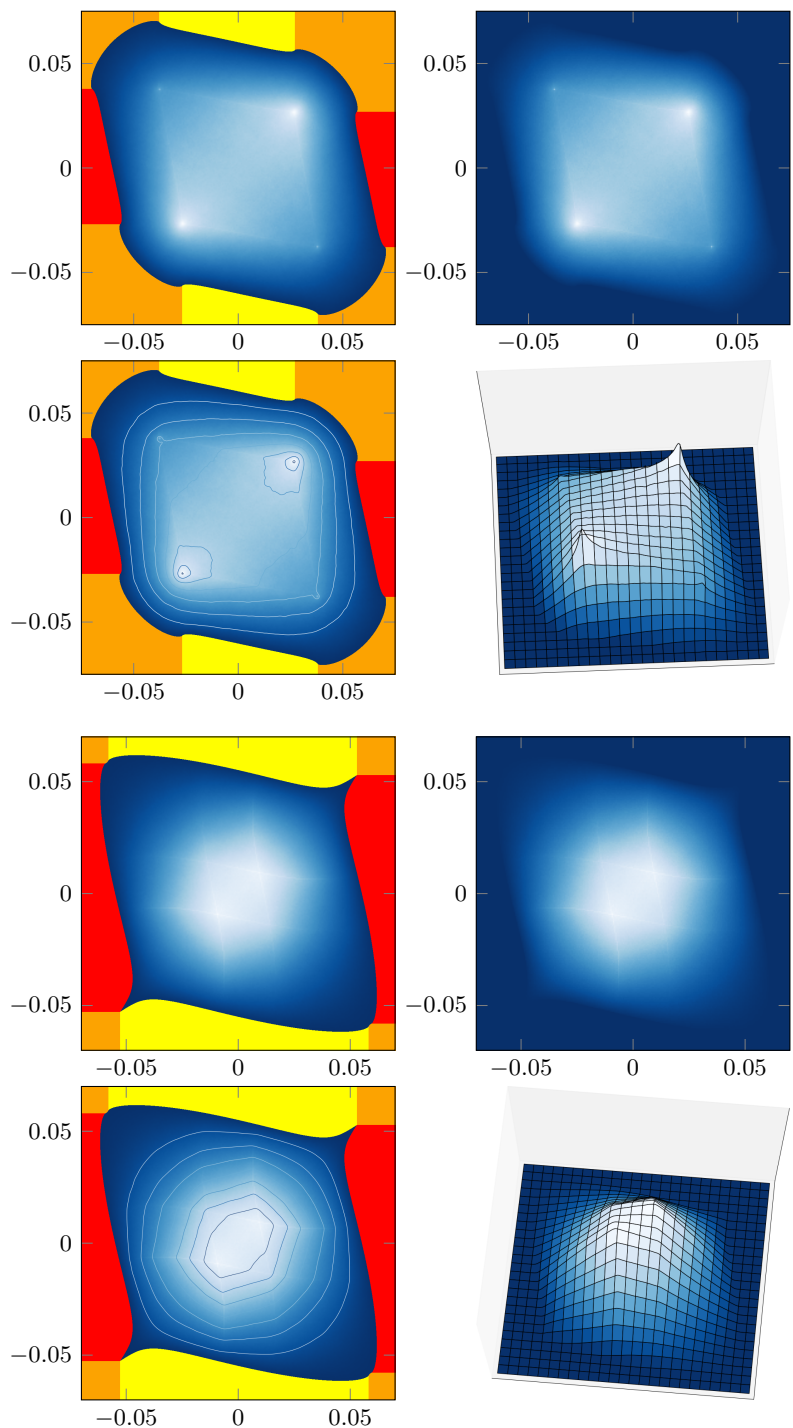

Figure 10: Long-term distribution of the optimal portfolio position, visualized by the probability density. These are shown as heatmaps, level sets, and as density surfaces.

| rel. pos. | $\beta$ | $\gamma$ | $r$ | $\mu_1$ | $\mu_2$ | $\sigma_1$ | $\sigma_2$ | $\rho$ | $\lambda_f$ | $\lambda_p$ | $z$ |
|---|---|---|---|---|---|---|---|---|---|---|---|
| Top | 1 | 3 | 0.03 | 0.08 | 0.08 | 0.4 | 0.4 | 0.30 | $ 1 | 3% | $ 10'000 |
| Bottom | 1 | 5 | 0.03 | 0.08 | 0.08 | 0.4 | 0.4 | 0.30 | $ 1 | 3% | $ 1'000 |